\documentclass{emulateapj}
\usepackage{wrapfig}
\usepackage{epsfig}
\usepackage{amsmath}
\usepackage{natbib}
\usepackage{array}

\newif\ifAMStwofonts
\AMStwofontstrue

\def\ga{\mathrel{\hbox{\rlap{\hbox{\lower4pt\hbox{$\sim$}}}\hbox{$>$}}}}
\def\la{\mathrel{\hbox{\rlap{\hbox{\lower4pt\hbox{$\sim$}}}\hbox{$<$}}}}

\newcommand{\NoAbsComps}{157}  
\newcommand{\NoOfSources}{31}  
\newcommand{\StackedSpinTemp}{$7200^{+1800}_{-1200}\rm\,K$} 
\newcommand{\NoHT}{11}

\newcommand{\NoLAB}{4}
\newcommand{\NoClaire}{16}
\newcommand{\NoObs}{53} 

\shorttitle{The 21-SPONGE Survey I}

\shortauthors{Murray et al.}
\begin{document}

\title{The 21-SPONGE H\textsc{i} Absorption Survey I: Techniques and Initial Results}

\author{Claire E. Murray$^1$$^{\dagger}$, Sne\v{z}ana Stanimirovi\'{c}$^1$$^{\dagger}$, W.~M.  Goss$^2$, John M. Dickey$^3$, Carl Heiles$^4$, Robert R. Lindner$^1$, Brian Babler$^1$, Nickolas M. Pingel$^1$, Allen Lawrence$^1$, Jacob Jencson$^{2,5}$, Patrick Hennebelle$^6$}
\affil{
$^1$Department of Astronomy, University of Wisconsin, Madison, 475 N Charter St, Madison, WI 53706, USA\\
$^2$National Radio Astronomy Observatory, P.O. Box O, 1003 Lopezville, Socorro, NM 87801, USA \\
$^3$University of Tasmania, School of Maths and Physics, Private Bag 37, Hobart, TAS 7001, Australia \\
$^4$Radio Astronomy Lab, UC Berkeley, 601 Campbell Hall, Berkeley, CA 94720, USA \\
$^5$Department of Astronomy, Ohio State University, 140 West 18th Avenue, Columbus, OH 43210, USA \\
$^6$Laboratoire AIM, Paris-Saclay, CEA/IRFU/SAp$-$CNRS$-$ Universit\'e Paris Diderot, 91191 Gif-sur-Yvette Cedex, France
}
\altaffiltext{$\dagger$}{cmurray@astro.wisc.edu, sstanimi@astro.wisc.edu}

\begin{abstract}
We present methods and results from ``21-cm Spectral Line Observations of Neutral Gas with the EVLA" (21-SPONGE), a large survey for Galactic neutral hydrogen (H\textsc{i}) absorption with the Karl G. Jansky Very Large Array (VLA). With the upgraded capabilities of the VLA, we reach median root-mean-square (RMS) noise in optical depth of $\sigma_{\tau}=9\times 10^{-4}$ per $0.42\rm\,km\,s^{-1}$ channel for the \NoOfSources{} sources presented here. Upon completion, 21-SPONGE will be the largest H\textsc{i} absorption survey with this high sensitivity. We discuss the observations and data reduction strategies, as well as line fitting techniques. We prove that the VLA bandpass is stable enough to detect broad, shallow lines associated with warm H\textsc{i}, and show that bandpass observations can be combined in time to reduce spectral noise. In combination with matching H\textsc{i} emission profiles from the Arecibo Observatory ($\sim3.5'$ angular resolution), we estimate excitation (or spin) temperatures ($T_s$) and column densities for Gaussian components fitted to sightlines along which we detect H\textsc{i} absorption (30/31). We measure temperatures up to $T_s\sim1500\rm\,K$ for individual lines, showing that we can probe the thermally unstable interstellar medium (ISM) directly. However, we detect fewer of these thermally unstable components than expected from previous observational studies. We probe a wide range in column density between $\sim10^{16}$ and $>10^{21}\rm\,cm^{-2}$ for individual H\textsc{i} clouds. In addition, we reproduce the trend between cold gas fraction and average $T_s$ found by synthetic observations of a hydrodynamic ISM simulation by Kim et al.\,(2014). Finally, we investigate methods for estimating H\textsc{i} $T_s$ and discuss their biases.
\end{abstract}


\section{Introduction}

Star formation drives the evolution of galaxies, and the material from which stars form constantly cycles in and out of the interstellar medium (ISM). Within this process, neutral hydrogen (H\textsc{i}) represents a crucial transitionary stage between supernovae-expelled plasma and dense star-forming gas. Consequently, the presence and influence of radiative and dynamic physical processes in the ISM are imprinted in the mass distribution of H\textsc{i} as a function of temperature and density. Observational constraints over the full range of temperature and density are therefore necessary for realistic heating, cooling and feedback prescriptions within accurate models of star formation and galaxy evolution on all galactic scales (e.g. Bryan et al.\,2007, Governato et al.\,2010, Ostriker et al.\,2010, Hill et al.\,2012, Christensen et al.\,2012). 

Analytical models of heating and cooling in the ISM predict the existence of two neutral H\textsc{i} phases-- the cold neutral medium (CNM) and the warm neutral medium (WNM)-- each in thermal equilibrium (e.g., Pikel'ner 1968, Field et al.\,1969, McKee \& Ostriker 1977). McKee \& Ostriker (1977) and Wolfire et al.\,(2003) predict that the two phases should be mostly in thermal equilibrium, with the CNM dominating and the WNM comprising only a few percent of the total H\textsc{i} column density. Over the range of pressures for which the two stable phases can coexist, $\sim1000<P/k<8000\rm\,cm^{-3}\,K$, we theoretically expect a volume density ($n$)  and kinetic temperature ($T_k$) of $(n,T_{\rm K})=(5$--$120\rm\,cm^{-3}$, 40--$200\rm\,K$) for the CNM, and $(n,T_{\rm K})=(0.03$--$1.3\rm\,cm^{-3}$, $4100$--$8800\rm\,K$) for the WNM \citep{Wolfire03}. Within the high-density CNM, collisions between electrons, ions and other H atoms can thermalize the 21-cm transition so that $T_s=T_k$. On the other hand, within the low-density WNM, collisions are not sufficient to thermalize the transition, and therefore it is widely expected that $T_s<T_k$ (e.g. Field et al.\,1958, Deguchi \& Watson 1985, Liszt 2001). 

In addition, the presence of H\textsc{i} in the intermediate, thermally unstable regime between the CNM and WNM has a range of theoretical implications. Dynamical processes are required to maintain this unstable material, as thermal pressure alone is not sufficient to force gas across extreme density boundaries (Cox et al.\,2005). Shocks driven by supernovae into turbulent, magnetized gas create dense CNM clouds within a multiphase medium, where the fraction of thermally unstable gas is determined by the supernova rate (Koyama \& Inutsuka 2002, MacLow et al.\,2005). Audit \& Hennebelle (2005) showed that turbulent flow collisions produce unstable fractions of 10-30\%, depending on the amount of turbulence injected. Modeling collisions of wind-blown superbubbles, Ntormousi et al.\,(2011) found most gas in cold clumps, with only 8-10\% thermally-unstable WNM by mass. In another colliding flow model, Clark et al.\,(2012) showed that the initial flow speed, including a time-dependent chemical network, has a huge effect on the resulting fractions of cold and warm gas. Clearly, observational constraints of the WNM properties and the full temperature distribution over $10$--$10^4\rm\,K$ are important to constrain the role of self-gravity, turbulence, and large-scale gas flows in models of the ISM, as well as to guide initial conditions for numerical simulations.

\subsection{Previous Work}

{
\setlength{\extrarowheight}{1pt}
\setlength{\tabcolsep}{2pt}
\begin{table*}
\caption{Survey Comparison}

\centering
\label{t:sens}
\begin{tabular}{l|cccccc}
\hline
\hline
\small
   &  Galactic & Angular res. & Velocity res. & Sensitivity & Number & Telescopes  \\
Survey & latitude ($^{\circ}$) &  (arcmin) & ($\rm km\,s^{-1}$) & ($\sigma_{\tau}$)$^a$ & of spectra & (absorption, emission) \\
 & (1) & (2) & (3) & (4) & (5) & (6) \\
\hline
{\bf Absorption targeted:} & & & & & & \\
\hline
21-SPONGE & $|b|>3.7$ & 4 & 0.5 & 0.0006 & 58 & VLA, Arecibo \\
Roy et al.\,2013 & $|b|>1.6$ & 36 & 1.0 & 0.0005 & 35 & WSRT/GMRT/ATCA, LAB \\
Heiles \& Troland 2003 & all $|b|$ & 4 & 0.16 & 0.002 & 78 & Arecibo \\
Stanimirovi\'c et al.\,2014 & $-32<b<-8$ & 4 & 0.16 & 0.002 & 27 & Arecibo \\
Mohan et al.\,2004 & $|b|>$15 & 36 & 3.3 & 0.005 & 102 & GMRT, LDS \\
\hline
{\bf Emission mapped:} & & & & & & \\
\hline
VGPS$^b$ & $|b|<$1.3 & 1 & 1.56 & 0.025-0.125 & 113 & VLA, GBT \\
CGPS$^c$ & -3.6$<b<$15 & 1 & 1.32 & 0.023-0.115 & 364 & DRAO, DRAO 26m \\
SGPS$^d$ & $|b|<$1.5 & 2 & 1.0 & 0.02-0.1 & 96 & ATCA, Parkes \\
GASKAP$^e$ & $|b|<$10 & 0.5 & 1.0 & 0.02 & $>$1000 & ASKAP \\
\hline
\end{tabular}
\label{t:sens}
\vskip 0.1 in
\footnotesize 
\raggedright
$^a$: Median RMS noise in H\textsc{i} optical depth per $1\rm\,km\,s^{-1}$ channel. \\
$^b$: \citet{VGPS};
$^c$: \citet{CGPS};
$^d$: \citet{SGPS};
$^e$: \citet{GASKAP}
\end{table*}
}

Observationally, the hyperfine 21-cm transition of H\textsc{i} provides an excellent tracer of the CNM and WNM. Given measurements of \emph{both} H\textsc{i} emission and H\textsc{i} absorption, it is possible to directly measure the excitation temperature (or spin temperature, $T_s$) and column density (N(H\textsc{i})) of CNM and WNM structures along the line of sight. Due to the high optical depth of the CNM, 21-cm absorption signatures are easy to detect, even with low sensitivity (e.g. Lazareff 1975, Dickey et al.\,1977, Crovisier et al.\,1978, Payne et al.\,1978, Dickey et al.\,1983, Braun \& Walterbos 1992, Heiles \& Troland 2003, Kanekar et al.\,2003, Mohan et al.\,2004, Roy et al. 2006, Begum et al.\,2010). Extensive studies have been undertaken in the Milky Way and the CNM is easily detected all the way out to a Galactocentric radius of $\sim25\rm\,kpc$ (e.g., Dickey et al.\,2009). 

However, the optical depth of the WNM is so low ($\tau\leq10^{-3}$) that extremely high sensitivity is required to detect it in absorption. Few studies have targeted individual detections of the WNM in absorption to directly measure $T_s$ (e.g. Carilli et al.\,1998, Dwarakanath et al.\,2002, Murray et al.\,2014), while many others estimate WNM $T_s$ from upper limits (i.e. $T_k$) or as line of sight averages in the presence of strongly absorbing CNM gas (Mebold et al.\,1982, Heiles \& Troland 2003, Kanekar et al.\,2003, Kanekar et al\,2011, Roy et al.\,2013a, Roy et al.\,2013b). Therefore, the mass distribution of H\textsc{i} in WNM-like conditions, i.e., at high $T_s$ and low $n$, is statistically unconstrained. 

Previous observational studies of CNM and WNM properties, although ubiquitous, typically do not reach sufficient sensitivity to detect the WNM or thermally unstable gas directly in H\textsc{i} absorption. The Millennium Arecibo 21 Centimeter Absorption-Line Survey (Heiles \& Troland 2003; HT03), comprised of 79 H\textsc{i} absorption and emission spectral pairs spread over the full Arecibo Observatory field of view, provided important constraints on the $T_s$ distribution of the CNM. A key result from HT03 is that that 48\% of the non-CNM spectral features had thermally unstable temperatures, with $500\leq T_s\leq 5000\rm\,K$. However, the high-$T_s$ measurements in the HT03 study are indirect (i.e., not based on direct detections of H\textsc{i} absorption). Furthermore, single dish telescopes have the disadvantage that emission fluctuations within the beam can contaminate absorption measurements, while interferometers resolve out this large scale structure. In addition, the HT03 results could be biased by low sensitivity. Increased integration time on non-detection sightlines from HT03 reveal absorption lines with $\tau<10^{-3}$, as shown by Stanimirovi\'{c} \& Heiles (2005). Previously-undetected, weak absorption lines with CNM-like $T_s$ will account for emission along the line of sight that was originally attributed to thermally unstable material, thereby reducing the estimated fraction of thermally unstable gas. Significant changes to the observed fraction of thermally unstable gas in the ISM will in turn have strong implications for the influence of any radiative and dynamic processes responsible for creating and maintaining the instability. 

In a study of H\textsc{i} absorption at the Karl G. Jansky Very Large Array (VLA), Begum et al.\,(2010) demonstrated that the VLA bandpass is stable enough to detect shallow (peak $\tau\sim10^{-3}$), wide (full width at half maximum (FWHM) $\sim7-8\rm\,km\,s^{-1}$) absorption lines. In combination with H\textsc{i} emission from the Arecibo Observatory, they identified individual absorption lines with spin temperatures in the thermally unstable regime, $T_s=400-900\rm\,K$, in observations towards 12 bright background sources. However, only $<30\%$ of their detected absorption lines had thermally unstable spin temperatures, which is much lower than the $48\%$ unstable fraction reported by HT03. This emphasized the need for a larger interferometric study of H\textsc{i} absorption at high sensitivity to further constrain the fractions of gas in all H\textsc{i} phases. 

Recently, Roy et al.\,(2013a,b) conducted an H\textsc{i} absorption survey of 35 continuum sources at the Westerbork Synthesis Radio Telescope (WSRT), Giant Metrewave Radio Telescope (GMRT) and Australia Telescope Compact Array (ATCA). Using the Leiden Argentine Bonn (LAB; Kalberla et al.\,2005) survey for H\textsc{i} emission, they found at least $28\%$ of the absorption-detected H\textsc{i} in the thermally unstable regime (Roy et al.\,2013b). However, this value is based on only 13 of their sources and the conservative assumptions that all detected CNM has $T_s=200\rm\,K$ and all non-detected WNM has $T_s=5000\rm\,K$, and not on direct measurements of the absorbing and emitting properties of the unstable gas.

\subsection{21-SPONGE}

To measure the physical properties of all neutral H\textsc{i} in the ISM, we are conducting a large statistical survey, ``21-cm Spectral Line Observations of Neutral Gas with the EVLA" (21-SPONGE), to obtain high-sensitivity Milky Way H\textsc{i} absorption spectra using the VLA. The recently upgraded capabilities of the VLA allow us to achieve median RMS noise levels in optical depth of $\sigma_{\tau}=9\times10^{-4}$ per 0.42 km\,s$^{-1}$ channel, which are among the most sensitive observations of H\textsc{i} absorption to date. Currently \NoOfSources{} sources are complete after over 200 hours of observing time. We have a very high detection rate so far, and we detect H\textsc{i} absorption in the direction of 30/\NoOfSources{} sightlines. 

In Murray et al.\,(2014) we applied a spectral stacking technique to the initial one third of the 21-SPONGE sample (19 H\textsc{i} absorption lines) and detected a widespread WNM component at high significance. The spin temperature of this component, equal to \StackedSpinTemp{} (68\% confidence), is higher than predictions based on collisional excitation alone ($<4400\rm\,K$, Liszt 2001), implying that additional excitation mechanisms must be prevalent in the ISM. Ly$\alpha$ scattering provides a likely source for this additional excitation. However, the required density of Ly$\alpha$ photons necessary to explain our result is significantly higher than expectations from recent theoretical and numerical studies (e.g. Liszt 2001, Kim et al.\,2014). For example, in their Galactic simulations, Kim et al.\,(2014) used a fixed Ly$\alpha$ photon number density of $10^{-6}\rm\,cm^{-3}$ and found $T_s=4000-5000\rm\,K$ for the WNM. Understanding the excitation processes of H\textsc{i}, as well as the strength and topology of the Ly$\alpha$ radiation field, clearly require future studies.

In Table~\ref{t:sens}, we compile properties of other single dish and interferometric surveys for H\textsc{i} absorption for comparison, including: (1) estimates of the area covered in Galactic latitude ($^{\circ}$), (2) the angular resolution in H\textsc{i} emission (arcmin), (3) the velocity resolution ($\rm km\,s^{-1}$), (4) the H\textsc{i} optical depth sensitivity ($\sigma_{\tau}$) per $1\rm\,km\,s^{-1}$ channels, (5) the number of H\textsc{i} absorption spectra, and (6) the telescopes used for H\textsc{i} absorption and emission. Those studies which targeted particular sources for measuring H\textsc{i} absorption at a desired sensitivity are classified as ``Absorption targeted", and those studies which mapped large spatial areas in H\textsc{i} emission and then extracted absorption from continuum sources within the map are classified as ``Emission mapped". We note that the 4 surveys in the latter category were all conducted at low-$|b|$ and therefore their spectra are more complex than those at high-latitude in the targeted absorption line surveys. 

21-SPONGE is one of the most sensitive and extensive absorption line surveys ever undertaken. We emphasize that our analysis will improve on previous work at similar sensitivity, as we estimate physical properties of interstellar H\textsc{i} using emission data obtained with spatial resolution closer to the $\sim1''$ interferometric absorption resolution from the VLA than previous studies. For example, Roy et al.\,(2013) use emission from the $\sim36'$-wide LAB beam, and we use emission data from the Arecibo Observatory, whose beam area ($\sim 3.5'\times3.5'$) is smaller than the LAB beam by two orders of magnitude. In addition, 21-SPONGE will occupy a unique space in terms of sensitivity and size for a long time, given that future H\textsc{i} absorption surveys with the Australian Square Kilometer Array Pathfinder (ASKAP) and the Square Kilometer Array (SKA) will not likely seek such high sensitivity in optical depth (e.g. McClure-Griffiths et al.\,2015).

{
\setlength{\extrarowheight}{1.2pt}
\begin{table*}
\caption{VLA Observation Information}

\centering
\label{tab:obs}
\begin{tabular}{lllcccccc}
\hline
\hline
\small
Source  & RA(J2000) & Dec (J2000) & l     & b     & $S_{1.4\rm\,GHz}$ &  $\sigma_{\tau}$    & Synthesized Beam & Time \\
  (name)      &  (hh:mm:ss)      &  (dd:mm:ss)    & ($^{\circ}$) & ($^{\circ}$) &  (Jy)$^a$      &   (10$^{-3}$)$^b$    & (arcsec$^2$)    & (hrs)$^c$ \\
\hline
4C32.44 & 13:26:16 & 31:54:10 & 67.240 & 81.049 & 4.6 & 1.5 &  2.8$\times$1.3 & 6.6\\
4C25.43 & 13:30:37 & 25:09:11 & 22.464 & 80.991 & 6.9 & 0.9 &  2.7$\times$1.2 & 2.9\\
3C286 & 13:31:08 & 30:30:33 & 56.526 & 80.676 & 14.9 & 0.7  &  4.1$\times$3.0 & 2.8\\
4C12.50 & 13:47:33 & 12:17:24 & 347.220 & 70.173 & 5 & 1.3  &  4.2$\times$1.5 & 2.8\\
3C273 & 12:29:06 & 02:03:05 & 289.945 & 64.358 & 54.9 & 0.6  &  7.8$\times$4.2 & 1\\
3C298 & 14:19:08 & 06:28:35 & 352.159 & 60.667 & 6 & 0.8     &  2.3$\times$1.4 & 2.8\\
4C04.51 & 15:21:14 & 04:30:19 & 7.290 & 47.748 & 4 & 2.9     & 12.2$\times$4.7 & 5.6\\
3C237 & 10:08:00 & 07:30:16 & 232.117 & 46.627 & 6.5 & 0.9   & 6.5$\times$4.4 & 5\\
3C225A & 09:42:15 & 13:45:51 & 219.866 & 44.025 & 4.4 & 1.5  &  3.9$\times$1.6 & 6\\
3C225B & 09:42:15 & 13:45:49 & 220.010 & 44.007 & 4.4 & 3.0  &  3.9$\times$1.6 & 6\\
3C345 & 16:42:59 & 39:48:37 & 63.455 & 40.948 & 7 & 1.3     &  3.4$\times$1.4 & 2.5\\
3C327.1 & 16:04:45 & 01:17:51 & 12.181 & 37.006 & 4.1 & 1.4  &  6.1$\times$4.8 & 4.7\\
3C147 & 05:42:36 & 49:51:07 & 161.686 & 10.298 & 22.9 & 0.4  & 4.4$\times$3.8 & 1.1\\
4C33.48 & 19:24:17 & 33:29:29 & 66.388 & 8.373 & 3.8 & 2.8  & 4.8$\times$1.8 & 5.7\\
3C154 & 06:13:50 & 26:04:36 & 185.594 & 4.006 & 5.2 & 0.9    &  14$\times$13 & 3.5\\
3C410 & 20:20:06 & 29:42:12 & 69.210 & -3.768 & 10 & 1.9    & 9.9$\times$5.5 & 2.5\\
B2050+36 & 20:52:52 & 36:35:35 & 78.858 & -5.124 & 5 & 2.3  &  4.3$\times$2.0 & 5\\
3C409 & 20:14:27 & 23:34:52 & 63.397 & -6.120 & 14 & 3.3    &  1.6$\times$1.4 & 3.3\\
PKS0531+19 & 05:34:44 & 19:27:21 & 186.761 & -7.109 & 7 & 0.5 &  1.3$\times$1.1 & 1.9\\
3C111 & 04:18:21 & 38:01:35 & 161.675 & -8.821 & 4.3 & 0.7   & 23$\times$23 & 5.1\\
3C133 & 05:02:58 & 25:16:24 & 177.725 & -9.914 & 5.7 & 1.8   & 1.3$\times$1.1 & 5.1\\
3C138 & 05:21:09 & 16:38:22 & 187.403 & -11.346 & 9 & 0.9    &  15$\times$5.2 & 2.2\\
3C123 & 04:37:04 & 29:40:13 & 170.581 & -11.662 & 47 & 0.7   & 19$\times$5.5 & 2.2\\
3C433 & 21:23:44 & 25:04:10 & 74.475 & -17.693 & 12 & 2.4   & 9.7$\times$5.7 & 3\\
3C120   & 04:33:11 & 05:21:15 & 190.373 & -27.397 & 3.4 & 1.1 & 4.7$\times$4.3 & 7\\
3C48 & 01:37:41 & 33:09:35 & 133.961 & -28.720 & 15.7 & 0.7  &  1.3$\times$1.2 & 2.8\\
4C16.09 & 03:18:57 & 16:28:32 & 166.633 & -33.598 & 8 & 0.8  & 1.3$\times$1.2 & 1.5\\
3C454.3 & 22:53:57 & 16:08:53 & 86.108 & -38.182 & 11 & 0.9 & 7.5$\times$3.9 & 2.5\\
J2232 & 22:32:36 & 11:43:50 & 77.436 & -38.582 & 7 & 0.7    &  5.2$\times$4.3 & 2.1\\
3C78  & 03:08:26 & 04:06:39 & 174.857 & -44.514 & 5.7 & 4.1   &  4.6$\times$3.7 & 5.2\\
3C459 & 23:16:35 & 04:05:17 & 83.038 & -51.285 & 4 & 0.8     & 5.7$\times$4.5 & 5\\
\hline
\end{tabular}
\vskip 0.1 in
\footnotesize 
\raggedright
$^a$: Condon et al.\,(1998). \\
$^b$: RMS noise in $\tau$ per 0.42 $\rm km\,s^{-1}$ channel, measured off-line. \\
$^c$: Total on-source time, not including calibration overheads.
\end{table*}
}

In this paper, we present the current progress of 21-SPONGE. In Section 2, we describe the observations and data reduction; in Sections 3 we discuss our methods of spectral line analysis via Gaussian fitting; in Secion 4 we show example decompositions for select sources; in Section 5 we calculate temperatures and column densities for all individual component fits; in Section 6 we compare our results with previous observational surveys; in Section 7 we compare our results with synthetic observations of a hydrodynamic Galaxy simulation by Kim et al.\,(2014); in Section 8 we discuss the pros, cons and biases of various methods for calculating $T_s$ from H\textsc{i} emission and absorption; and in Section 9, we summarize the results.

\section{Data Processing}

\subsection{Observations}
\label{s:obs}

VLA observations for 21-SPONGE began in February 2011, and targeted 58 sources from the NRAO/VLA Sky Survey (NVSS) catalog with flux densities at 1.4 GHz ($S_{1.4\rm\,GHz}) >3\rm\,Jy$ (Condon et al. 1998). We chose most sources to lie at high galactic latitude ($|b|$ $>$ 10$^\circ$) to avoid complicated H\textsc{i} profiles associated with the Galactic plane, and also to have angular sizes less than $1\,\rm '$ to avoid resolving substantial continuum flux. The range of source parameters allows us to conduct observations in all array configurations and configuration-moves except for D-array (most compact configuration). Currently, \NoObs{} sources have received full integration time and \NoOfSources{} of these \NoObs{} have been fully reduced and analyzed. Table~\ref{tab:obs} contains information about the completed observations. 

All VLA observations use simultaneously three separate, standard L-band configurations, each with one dual-polarization IF of bandwidth 500 kHz and 1.95 kHz per channel spacing. One IF is centered at $1.420408\,\rm GHz$ (H\textsc{i} line rest frequency, ``standard"), one at $1.421908\,\rm GHz$ ($1.5\,\rm MHz$, or about $300\rm\,km\,s^{-1}$, higher than the H\textsc{i} rest frequency, called ``high") and one at $1.418908\,\rm GHz$ ($1.5\,\rm MHz$ lower, called ``low"). We use the high and low configurations for observing our bandpass calibrators to avoid contamination by local H\textsc{i} in the direction of all calibrators at the H\textsc{i} rest frequency. Because we normalize our solution with respect to the continuum level, the absolute phase change associated with the frequency switching method is not an issue. We use the ``standard" configuration for observing target sources and measuring H\textsc{i} absorption, allowing for a velocity coverage of $107.5\,\rm km s^{-1}$ with $0.42\rm\,km\,s^{-1}$ channels, corresponding to a velocity resolution of $0.5\rm \,km\,s^{-1}$ (Rohlfs \& Wilson 2004). Relative flux calibration is performed via self-calibration on the target source and phase and amplitude calibration are performed by observing a nearby VLA calibrator source. 

For bandpass calibration, we observe the strong calibrators 3C286, 3C48 or 3C147 in both high and low instrument configurations following each on-source scan. For each full-length observing session (between 3 and 5 hours) we allocate up to 80\% overhead for the high and low bandpass observations. However, if the source was stronger than $\sim7\,\rm Jy$, it was observed as its own bandpass calibrator to conserve slewing time over the course of the observation, which is particularly beneficial for filler-project length observing files (maximum 1 hour total).

\subsection{Data reduction}

All data sets were reduced using the Astronomical Image Processing System\footnote{http://www.aips.nrao.edu} (AIPS). We decided to use AIPS instead of CASA\footnote{http://casa.nrao.edu/} to take advantage of the suite of bandpass analysis tasks unique to AIPS when we began the survey. During observations conducted in configuration moves involving D-array, all baselines shorter than $300\,\rm m$ were excluded to avoid contribution from partially resolved H\textsc{i} emission. After initial flagging of all observations, the task BPLOT was used to examine initial bandpass solutions for the high and low bandpass calibration observations in detail for each antenna. These solutions were then combined to produce a final bandpass solution. For all sources, we consistently reach noise levels in normalized bandpass amplitude profiles of $<10^{-3}$ per $0.42\rm\,km\,s^{-1}$ channels.  

In addition to a linear slope in the bandpass solution, we found a stable, periodic ripple which was revealed only because of the extreme sensitivity of our observations. The ripple is caused by the finite impulse response (FIR) filters applied to the data prior to correlation, and is stable in amplitude, phase and time. Appendix A contains information on the properties and stability of the bandpass structure. 

Furthermore, we demonstrated that the VLA bandpass is stable enough to combine bandpass observations conducted at different times or on different days. This is especially helpful for filler observations (max total observation time $\leq$1 hour). We are able to combine bandpass observations between separate days, thereby increasing the total bandpass observation time. The noise level in the bandpass solution decreases with added time by index of $-0.44\pm0.02$, which is nearly the theoretically expected index of $-0.5$ (i.e. by $1/\sqrt{t_2/t_1}$ for combined integration time $t_2$ and initial integration time $t_1$). Appendix A contains further details on the noise improvement due to time averaging in the cases of additional datasets.

Following amplitude and phase calibration on the target dataset, we apply the bandpass solution from the combined bandpass calibration observation. We then perform self-calibration on the target dataset by isolating the target source continuum, constructing an image, and using it to calibrate the continuum dataset. After applying the self-calibration solution, we repeat the process until the signal-to-noise level in the resulting continuum image no longer improves significantly. For most sources, this occurs after only two iterations. 
\begin{figure}[b!]
\begin{center}
\vspace{-15pt}
\includegraphics[width=0.49\textwidth]{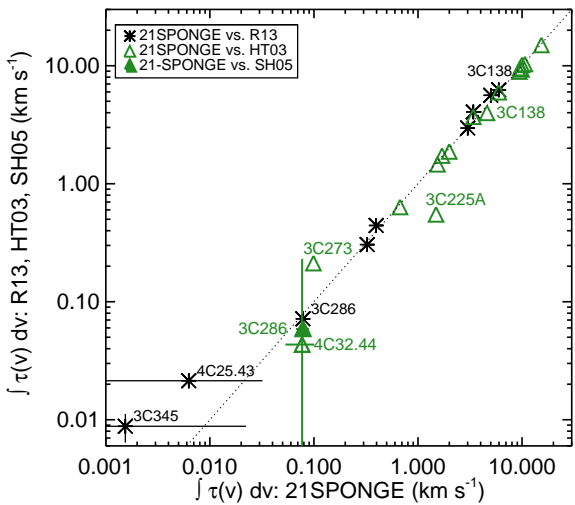}
\vspace{-100pt}
\caption{
Comparing integrated optical depth values between 21-SPONGE and Roy et al.\,2013 (R13; black stars), 21-SPONGE and HT03 (green diamonds) and 21-SPONGE and Stanimirovi\'c \& Heiles (2005) (SH05; filled green diamond). The dotted line shows y=x to highlight the agreement between the majority of points. Sources which strongly deviate from y=x are indicated by their source names. Errors on the measurements are included, but are smaller than the symbol sizes unless visible.
}
\vspace{-2pt}
\label{f:tau}
\end{center}
\end{figure}
Returning to the original target dataset, the continuum contribution to each source is determined by fitting a linear model to line-free channels. We subtract this from the source visibilities in all channels using the AIPS task UVLSF. We then use the task CVEL to correct source velocities for Earth's rotation and motion towards the Local Standard of Rest (LSR) within the Solar System.\footnote{We initially applied CVEL before continuum subtraction (UVLSF). However, we observed considerable Gibbs ringing in the edge channels of our reduced spectra. This forced us to Hanning smooth all datasets, reducing our velocity resolution by a factor of 2. After investigating the origin of the Gibbs ringing, we found that the ringing phenomenon arose in the order of CVEL and continuum subtraction, and reversing their order of application eliminated the ringing effect.}

The final calibrated data cubes are constructed using the AIPS task IMAGR, and are cleaned to three times the background noise in a test image of one channel. For each cube, the pixel sizes are calculated to be 4 times smaller than the beam size to properly sample the beam. Finally, we extract the absorption spectrum from the central pixel of each image, as in most cases, the sources are unresolved by the observations. However, of the \NoOfSources{} completed sources, 5 were resolved (3C327.1, 3C154, 3C111, 3C133, 3C123). To extract the spectra from these 5 data cubes, we average all spectra over the extent of the source. 

The achieved sensitivities for the final data products are listed in Table~\ref{tab:obs}, and the quoted values for $\sigma_{\tau}$ are per $0.42\rm\,km\,s^{-1}$ channel. The median RMS noise values are $\sigma_{\tau}=9\times10^{-4}$ (mean $1\times10^{-3}$) per $0.42\rm\,km\,s^{-1}$ channel, and $\sigma_{\tau}=6\times10^{-4}$ (mean $8\times10^{-4}$) per $1\rm\,km\,s^{-1}$ channel. Several sources have high noise due to the fact that they were resolved by the VLA, and we plan to include additional data in these directions to reduce the noise for the final data release. 

\subsection{Optical Depth Comparison}

To verify the accuracy of our absorption spectra, we compare integrated optical depths with two previously published surveys: the Arecibo Millennium Survey (HT03) composed of 79 absorption spectra with matching expected emission spectra from the Arecibo Observatory, and 35 high-sensitivity absorption spectra from the GMRT, WSRT and ATCA by Roy et al.\,(2013). We compare the sources overlapping between 21-SPONGE and HT03 (green diamonds) and between 21-SPONGE and Roy et al.\,(2013) (black stars) in Figure~\ref{f:tau}.

The three catalogs generally agree well in terms of integrated optical depth. In Figure~\ref{f:tau}, we indicate the source names of those which differ most between the catalogs. In the cases of 4C25.43 and 3C345, and Roy et al.\,(2013) have better sensitivity ($\sigma_{\tau}=5\times10^{-4}$ and $5\times10^{-4}$ whereas we achieve $\sigma_{\tau}=13\times10^{-4}$ and $9\times10^{-4}$ respectively), so our $\int \tau(v) dv$ values have larger error bars. For source 3C273, the HT03 baseline is not perfectly flat, with the consequence that their $\int \tau(v) dv$ value is higher than ours. HT03 did not measure any H\textsc{i} absorption above the $1\sigma$ limit towards 3C286, so we compare our optical depth spectrum to the higher-sensitivity result from Stanimirovi\'c \& Heiles (2005; SH05) (filled green diamond), and observe that our integrated values are consistent. In addition, we agree very well with the Roy et al.\,(2013) value for 3C286. Towards source 4C32.44, HT03 have lower sensitivity and so their value has larger error bars. Finally, the HT03 profile towards 3C225A, a resolved double-lobed radio galaxy, contains ``positive absorption" due to an artifact caused by the H\textsc{i} emission fluctuations and large uncertainties in estimating H\textsc{i} emission profiles.

Remaining residual differences that exist could possibly be caused by fluctuations in absorption profiles due to transient clouds, as has been well-studied in the cases of 3C147 (Lazio et al. 2009) and 3C138 (Brogan et al. 2005, Roy et al. 2012). These studies find differences in optical depth between observational epochs (spaced by several years) of up to $\sim0.5$. 

Overall, the excellent agreement between 21-SPONGE absorption and HT03, Roy et al.\,(2013) and Stanimirovi\'c \& Heiles (2005) absorption demonstrates that our observing and reduction strategies are sound. It is also encouraging to see that in the majority of cases, 21-SPONGE observations agree well with HT03 and Stanimirovi\'c \& Heiles (2005). This proves that Arecibo - a single dish telescope - can reliably produce accurate H\textsc{i} absorption profiles in the directions of strong sources ($S_{\rm1.4\,GHz}>3\rm\,Jy$).

\subsection{Matching Emission Profiles}

We obtained matching H\textsc{i} emission profiles along most sightlines from the 305-m Arecibo Observatory as part of project a2770. For the observations, we used the L-wide receiver to simultaneously observe H\textsc{i} at 1420 MHz and three OH lines (1665, 1667 and 1720 MHz) with two linear polarizations. We achieve a velocity resolution of $0.16\rm\,km\,s^{-1}$ over 2048 channels between $-164$ and $164\rm\,km\,s^{-1}$ for the H\textsc{i} observations. The $\sim$3.5' angular resolution of the Arecibo telescope at these frequencies complements the VLA observations by minimizing the effects of mismatched beam sizes on interpreting the absorption and emission spectra. It would require prohibitively long integration times to acquire emission profiles at similar sensitivity from the VLA in D array. 

Following the methods of HT03, we constructed an ``expected" H\textsc{i} emission profile ($T_{\mathrm{exp}}(v)$) toward each source by interpolation using a pattern of 16 observed off-source positions. The expected profiles represent the profile one would observe if the continuum source were turned off. The profile construction via least squares fit is described fully by HT03, and was also implemented by Stanimirovi\'c \& Heiles (2005) and Stanimirovi\'c et al.\,(2014). The error in $T_{\mathrm{exp}}(v)$ as a function of velocity, $\sigma_{T_{\rm exp}}(v)$, is also computed in this process based on the difference in system temperature and spatial offset between on-source and off-source pointings (HT03, Section 2.7). As discussed by Stanimirovi\'c et al.\,(2014), we used a simpler method than HT03 and excluded fine tuning for gain variations under the assumption that correctly accounting for them requires accurate knowledge of the spatially-varying gain and beam shape. In addition, we used a second-order Taylor expansion to construct $T_{\mathrm{exp}}(v)$, which, as found by Stanimirovi\'c et al.\,(2014), is a noisier but more accurate approach than using a first-order Taylor expansion. We then divided the expected profiles by a total beam efficiency factor of 0.85 (P. Perillat private communication) to convert the expected profiles to brightness temperature. 

To compare our results, we scaled the expected emission profiles from HT03 to our new Arecibo expected emission profiles by their velocity integrals and found that applying a median beam efficiency factor of 0.81 to the HT03 profiles produced the best agreement with our beam efficiency-scaled profiles from a2770. For each source, we then selected the scaled emission profile with the best sensitivity. For \NoHT{}/\NoOfSources{} sources we use the HT03 expected emission profiles, and for \NoClaire{}/\NoOfSources{} sources we use a2770 data. The remaining \NoLAB{} sources do not have Arecibo emission profiles because they are on the edge of the Arecibo field of view, and so we use emission spectra from the Leiden Argentine Bonn (LAB) survey for these sources (Kalberla et al.\,2005). 

\subsection{Stray Radiation}

Due to the complex beam pattern of the Arecibo telescope, our emission profiles are likely cotaminated at some level by radiation from higher order side-lobes, known as stray radiation. To estimate the level of this contamination, we use additional data from the Galactic Arecibo L-band Feed Array Survey in H\textsc{i} (GALFA-H\textsc{i}; Stanimirovi\'c et al.\,2006, Peek et al.\,2011). GALFA-H\textsc{i} is an all-sky survey for H\textsc{i} emission in the Galaxy using the seven beam-array ALFA receiver. Peek et al.\,(2011) smoothed GALFA-H\textsc{i} data cubes to the $\sim36'$ angular resolution of the stray radiation-corrected LAB survey (Kalberla et al.\,2005) and found that the differences generally fall within typical $1\sigma$ errors (see Figure 12 in Peek et al.\,2011). Stray radiation signatures appear as broad, wing-like features at high velocities, which are easy to isolate visually from the narrower, peak-like differences due to mismatched beam shapes between the LAB survey and the smoothed GALFA-H\textsc{i} data cubes. 

To compare our results for all sources covered by GALFA-H\textsc{i} to date, we constructed expected GALFA-H\textsc{i} emission profiles around each source by averaging $5\times5$ pixel$^2$ regions around each source. The central pixels of this average are likely affected by absorption, so we removed the central $3\times3$ pixel$^2$ region which corresponds to roughly one beam area. We then subtracted our $T_{\mathrm{exp}}(v)$ profiles from the expected GALFA-H\textsc{i} profiles and found that the differences fall mostly within $3\sigma$ errors in $T_{\mathrm{exp}}(v)$, which is very encouraging. The best method of removing the stray radiation contamination in lieu of modeling the complex, zenith-angle-dependent beam pattern of the Arecibo telescope would be to bootstrap the GALFA-H\textsc{i} data to the LAB survey and simply remove the differences from our profiles (e.g. as discussed by Peek et al.\,2011). However, applying this method necessarily injects noise from the LAB survey into the higher-sensitivity GALFA-H\textsc{i} profiles. In addition, the method assumes that the contamination to GALFA-H\textsc{i} (observed with ALFA) is the same as to our $T_{\mathrm{exp}}(v)$ profiles (observed with L-wide), which may not necessarily be the case. In the future we plan to investigate alternative methods for modeling the stray radiation contribution to our emission profiles without degrading our sensitivity unnecessarily. 

\section{Analysis}

\subsection{Gaussian Fitting}
\label{sec_gauss}

To estimate physical properties of individual interstellar clouds along the line of sight, we decompose the H\textsc{i} spectra into Gaussian functions (e.g. HT03). Although the CNM dominates the absorption spectra, by achieving extremely high sensitivities we aim to detect shallow, wide signatures of warm gas directly in absorption. 

Based on the two-phase ISM model, discussed in detail by \citet{Dickey03} and HT03, both CNM and WNM contribute to the expected profile, $T_{\rm exp}(v)$. We denote the CNM and WNM components detected in absorption by the subscript ``ABS" and the WNM components detected in emission by the subscript ``EM". Thus $T_{\rm exp}(v)$ is given by,
\begin{equation}
T_{\rm exp}(v)= T_{B,\mathrm{ABS}}(v) + T_{B,\mathrm{EM}}(v),
\label{e:texp}
\end{equation}
where $T_{B,\rm ABS}(v)$ is the brightness temperature due to absorption-detected gas, and $T_{B,\rm EM}(v)$ is the brightness temperature due to non-absorbing gas detected only in emission.

For each absorption spectrum, we identify a number ($N$) of Gaussian components using
a least-squares fit (described in Section 3.3), so that the absorption spectrum $\tau(v)$ is given by,
\begin{equation}
\tau(v) = \sum_{n=0}^{N-1} \tau_{0,n} e^{-[(v-v_{0,n})/\delta v_n]^2},
\end{equation}
where ($\tau_0,v_{0,n},\delta v_n$) are the (peak optical depth, central velocity, 1/$e$ width) of the $n^{\rm th}$ component
in the fit. 

Following the fit to the absorption spectrum, and assuming each fit component is independent and 
isothermal with spin temperature $T_{s,n}$, the brightness temperature contribution from the $N$ Gaussian functions detected in absorption to the expected emission profile is given by,
\begin{equation}
T_{B,\mathrm{ABS}}(v) = \sum_{n=0}^{N-1} T_{s,n} (1-e^{-\tau_n(v)})e^{-\sum_{m=0}^{M-1}\tau_m(v)}.
\label{tb:cnm}
\end{equation}
Here, the subscript $m$ and the associated optical depth $\tau_m(v)$ run through the $M$ number of
CNM clouds lying in front of cloud $n$. 
\begin{figure*}[t!]
\begin{center}
\label{figa}\includegraphics[width=0.32\textwidth]{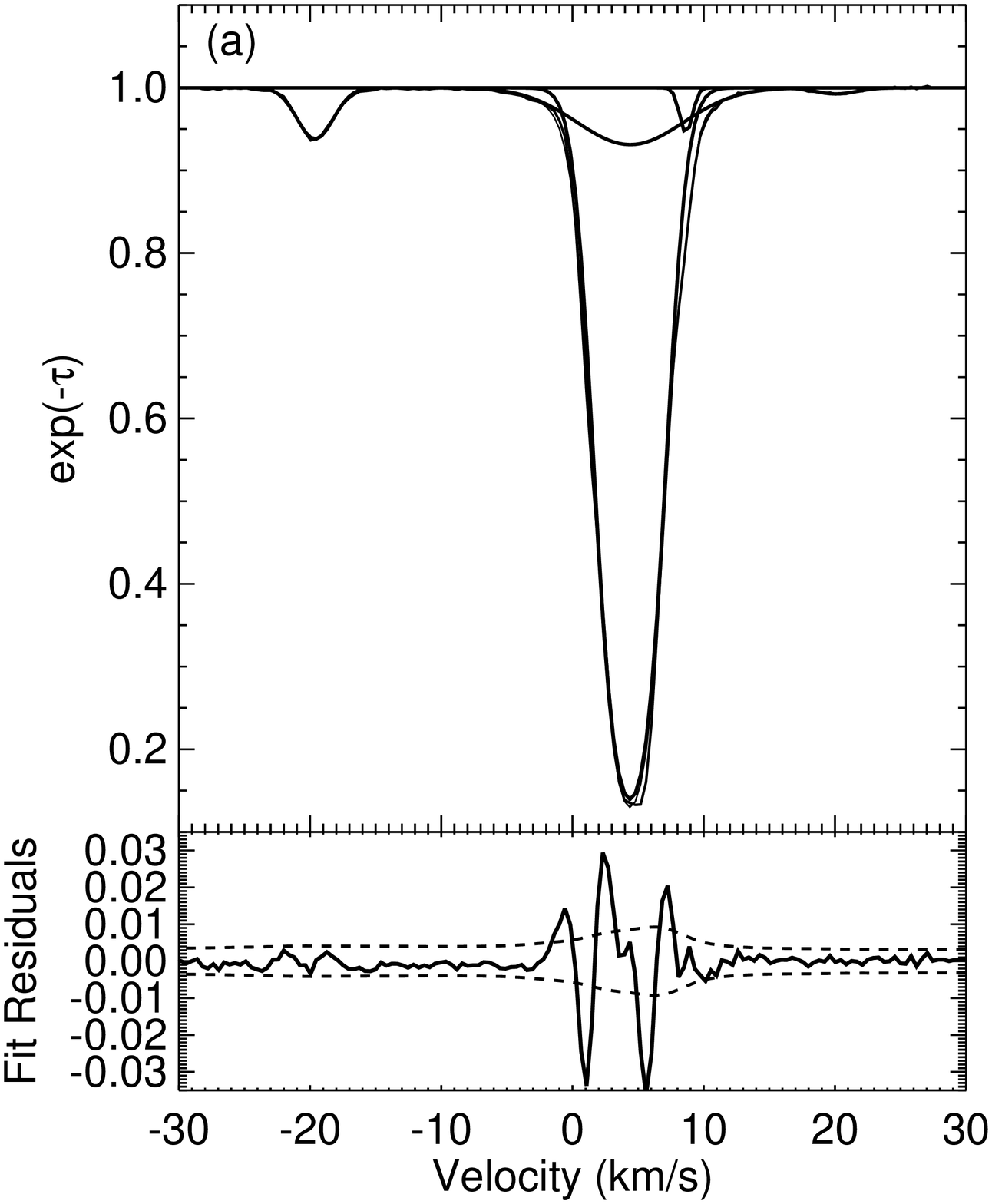}
\label{figb}\includegraphics[width=0.32\textwidth]{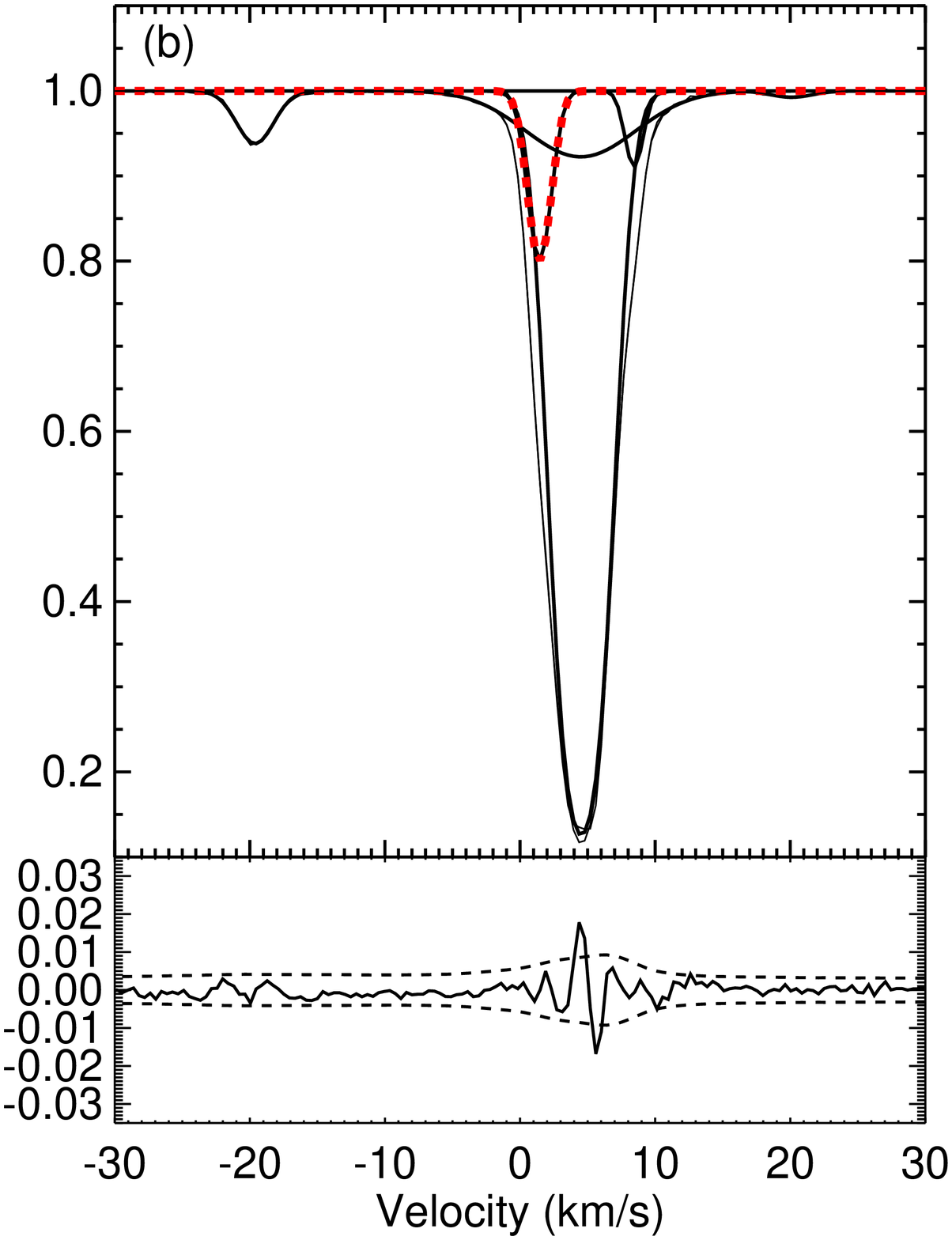}
\label{figc}\includegraphics[width=0.32\textwidth]{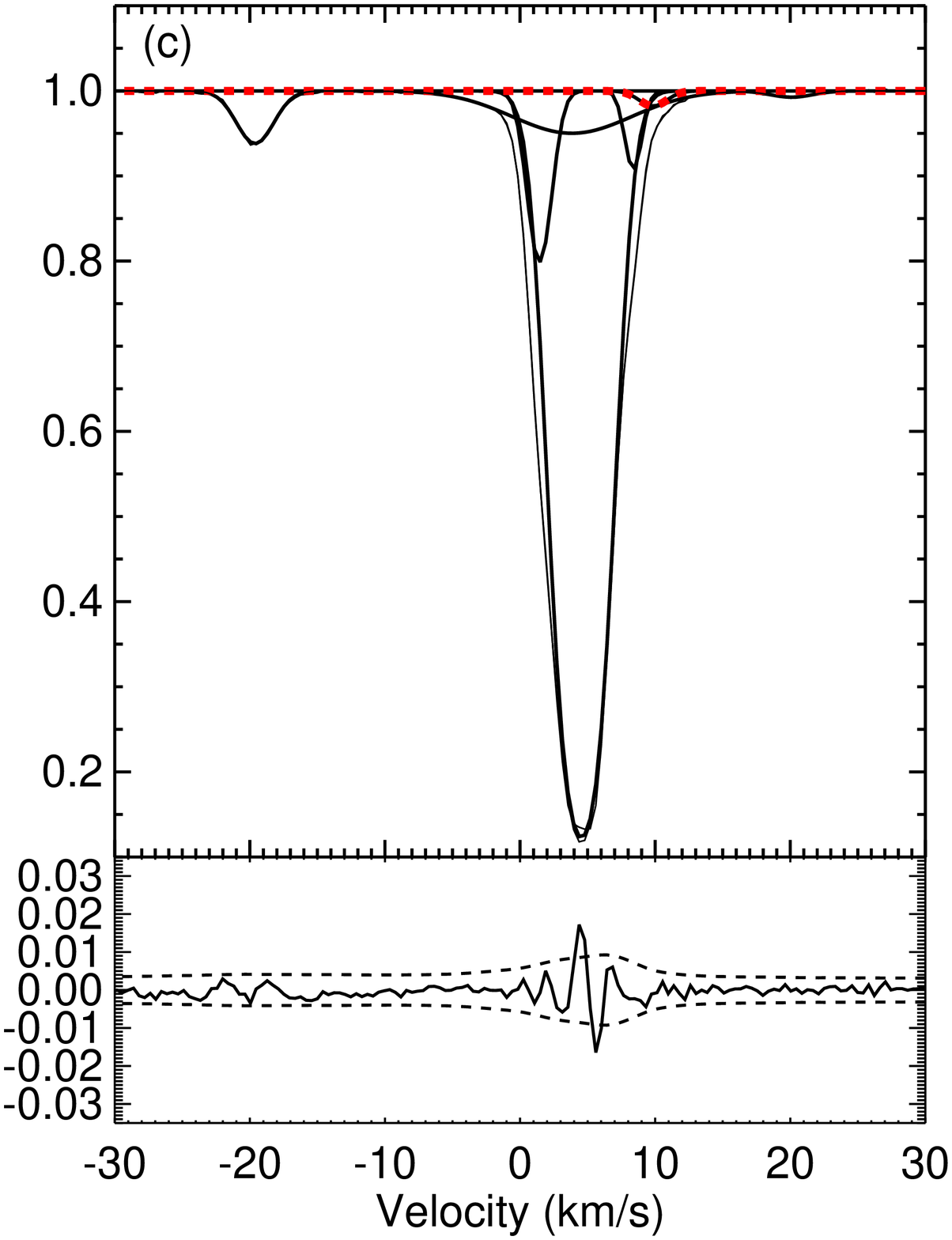}
\vspace{0pt}
\caption{
Three Gaussian fits to the optical depth profile in the direction of 3C123 demonstrating the fit improvement with additional components. Top panels: absorption spectrum (exp($-\tau$)) with Gaussian components overlaid; Bottom panels: fit residuals with $\pm\sigma_{\tau}(v)$ overlaid (see Section~\ref{s:noise}). (a): base fit, 5 components, (b): additional component added (thick red dashed), 6 total, improving the fit with 99\% confidence by the F-test, so the component is retained, (c): additional component added (thick red dashed), 7 total, improving the fit with 61\% confidence by the F-test, so the additional component is rejected. 
}
\vspace{-10pt}
\label{3c123}
\end{center}
\end{figure*}

\begin{figure*}
\begin{center}
\label{figa}\includegraphics[width=0.49\textwidth]{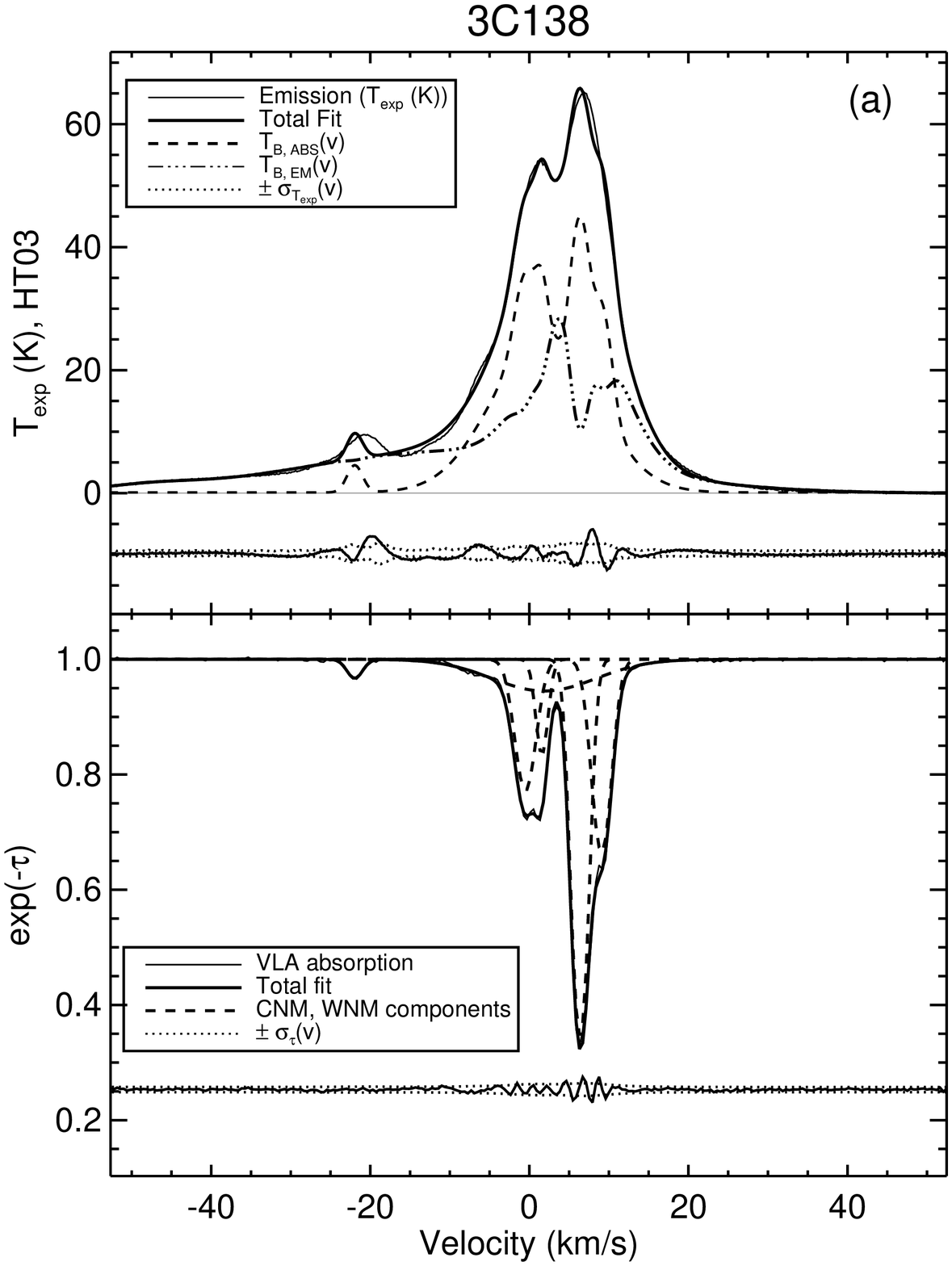}
\label{figb}\includegraphics[width=0.49\textwidth]{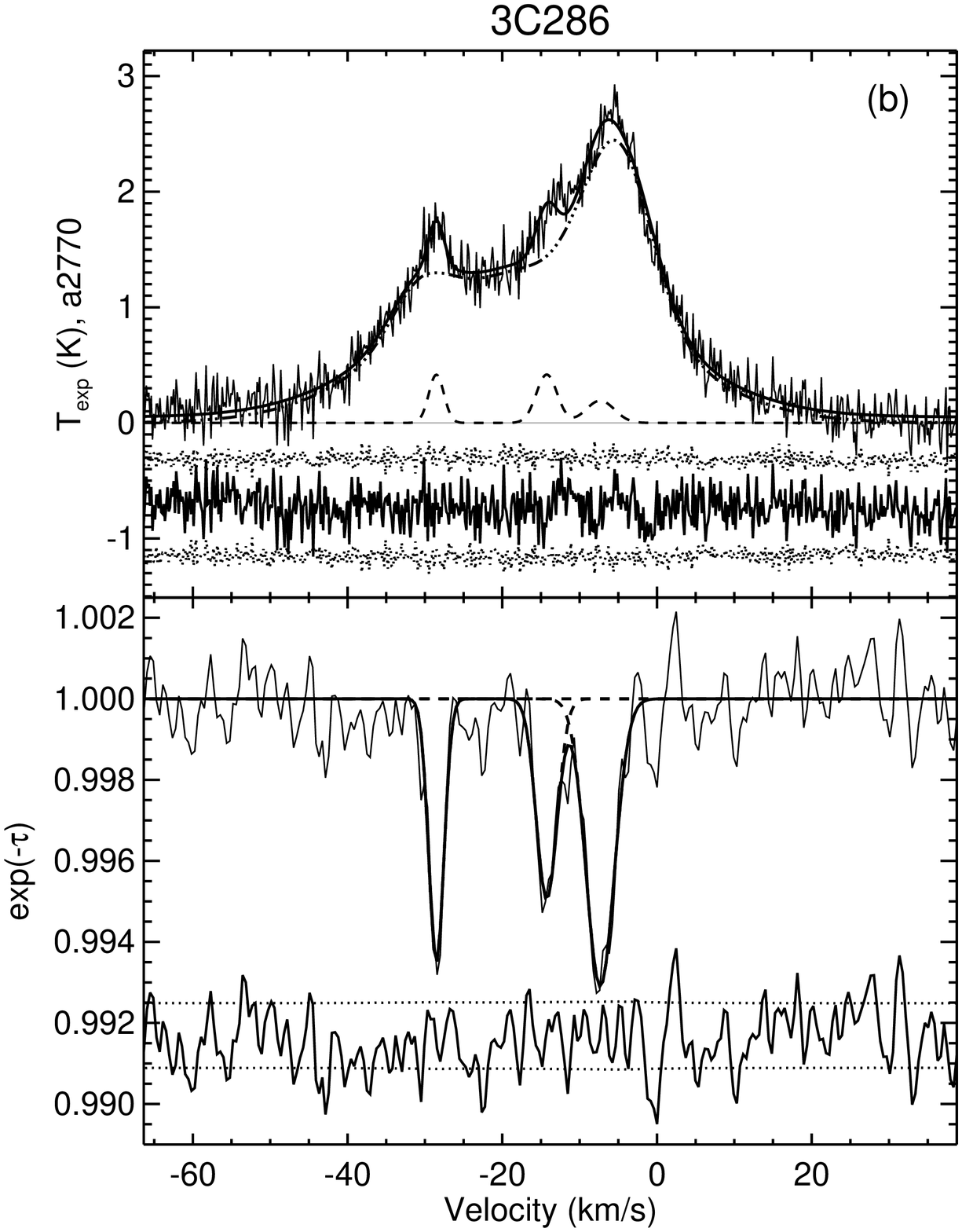}
\label{figc}\includegraphics[width=0.49\textwidth]{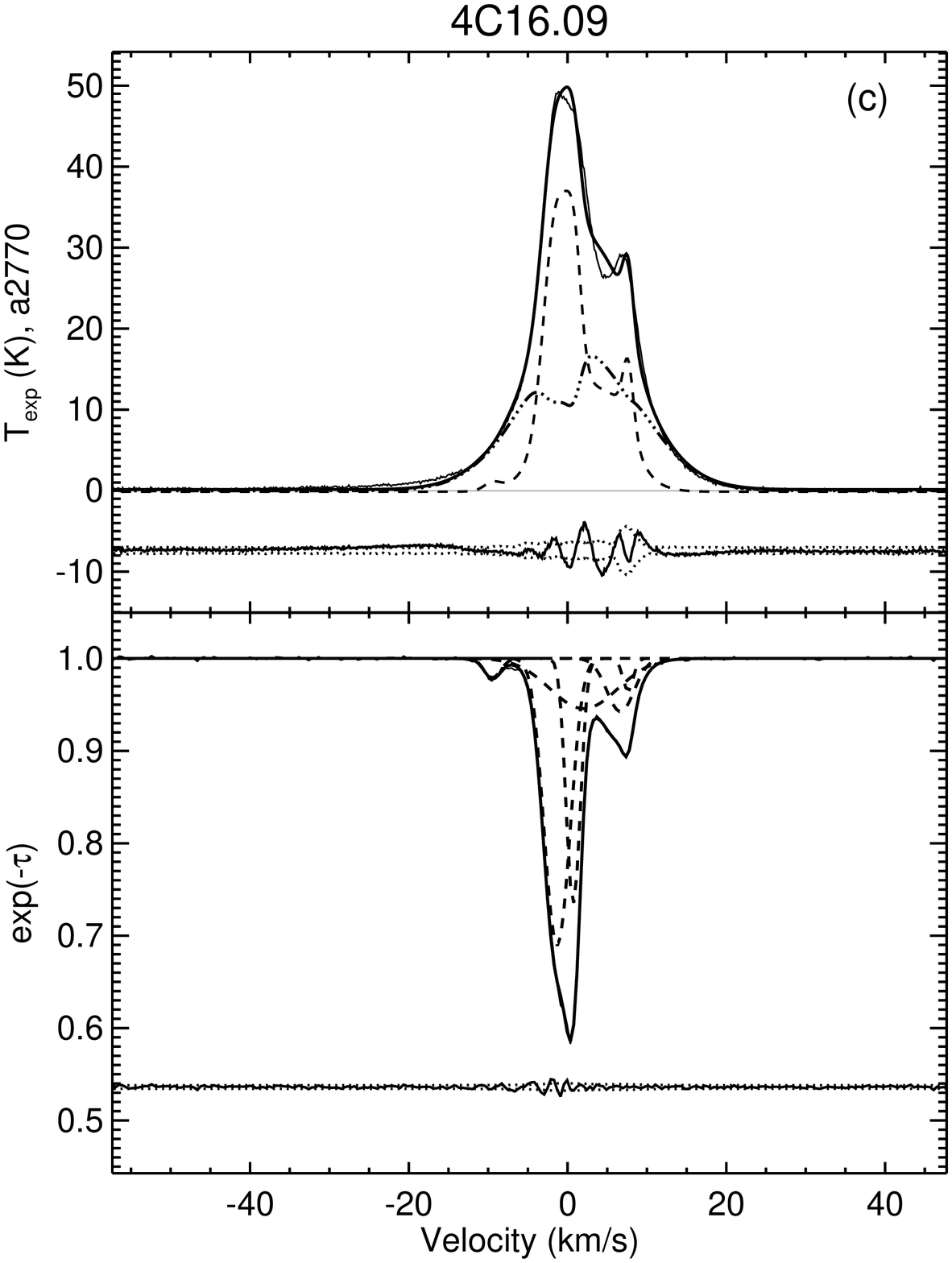}
\label{figd}\includegraphics[width=0.49\textwidth]{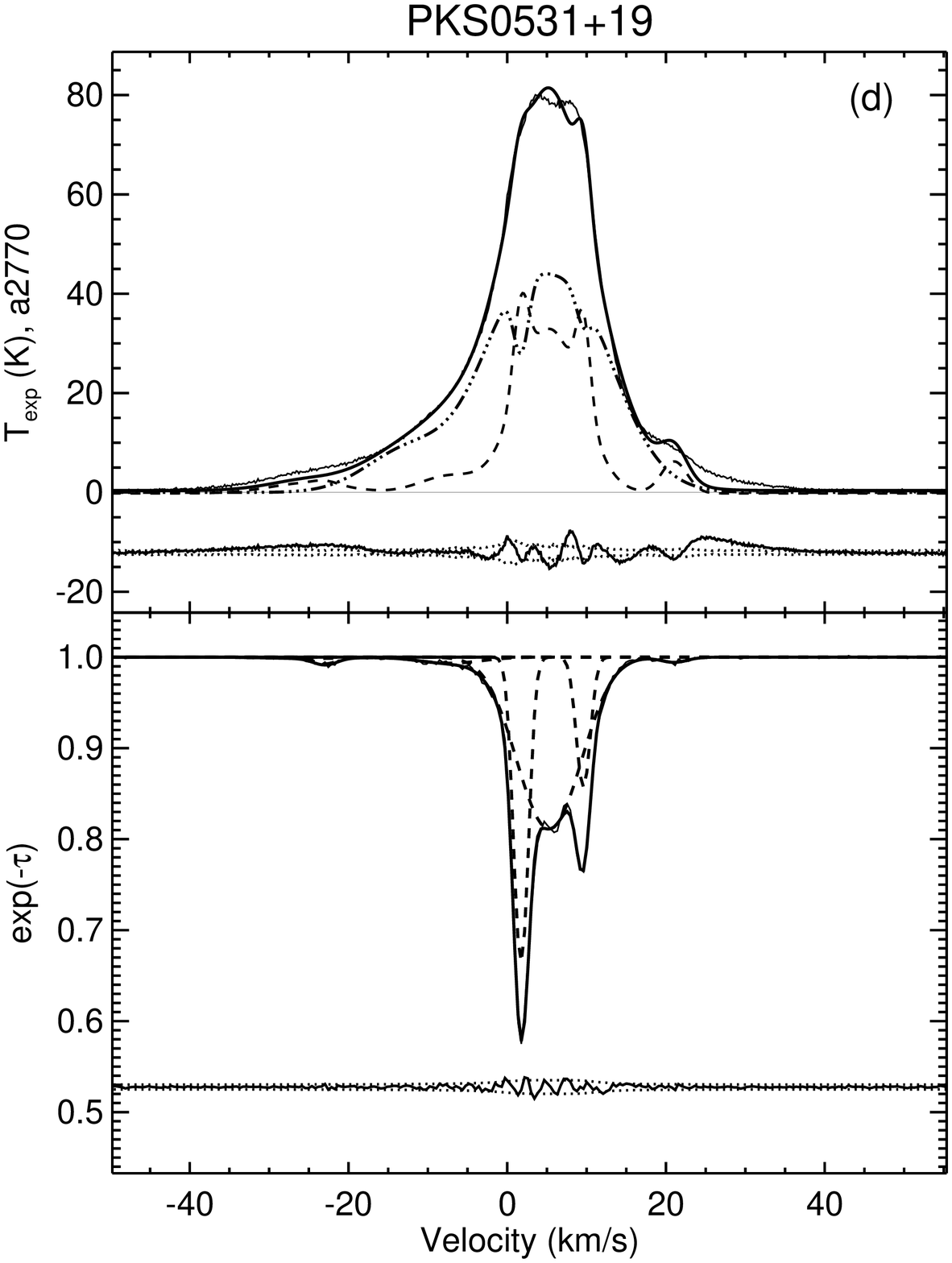}
\caption{\label{f:examples}
Gaussian fitting decomposition for 4/\NoOfSources{} completed sources. The sources are displayed in two-panel plots, where the top panel displays the emission profile (from HT03, a2770, or the LAB survey) and the bottom panel displays the VLA absorption profile. The fits are overlaid in several forms, including total CNM and WNM contribution and individual components, according to the inset legend in panel (a). Residuals in the fits are offset at the bottom of each panel, with $\pm\sigma_{T_{\rm exp}}(v)$ (top, see Section 2.4) and $\pm\sigma_{\tau}(v)$ (bottom, see Section 3.2) to illustrate the goodness of fit. 
}

\label{f:selres}
\end{center}
\end{figure*}

\begin{figure}
\begin{center}
\includegraphics[width=0.5\textwidth]{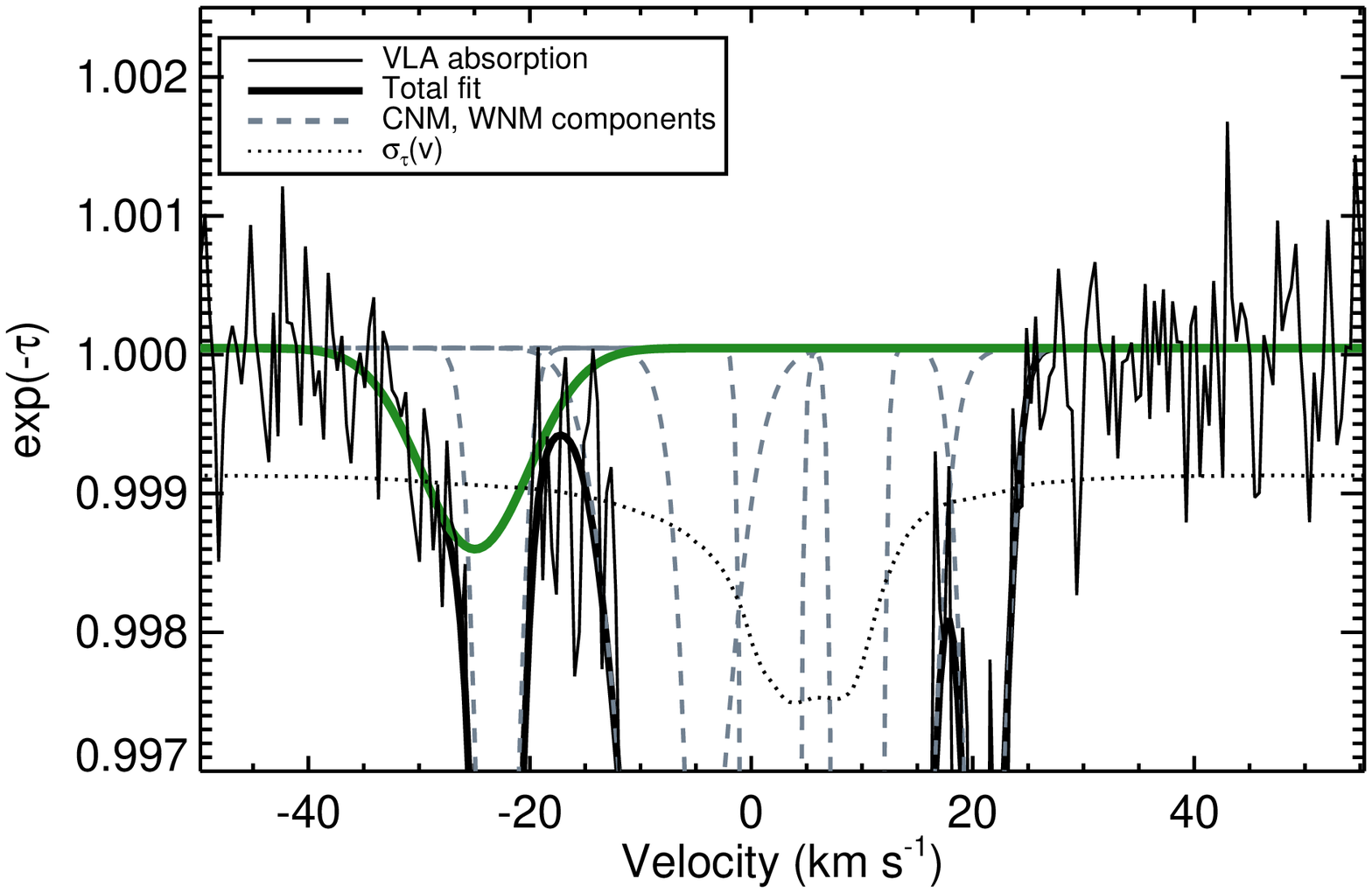}
\vspace{-120pt}
\caption{\label{f:pks_zoom}
Zoom-in of a broad, warm component (green) in the complex profile towards PKS0531+19. Velocity: $v_{0,\rm LSR}=-24.96\pm4.25\rm\,km\,s^{-1}$, FWHM: $\Delta v=11.5\pm9.3\rm\,km\,s^{-1}$, peak optical depth: $\tau_{\rm peak}=0.15\pm0.17$, maximum kinetic temperature: $T_{k,\rm max}=2900\pm5000\rm\,K$, spin temperature: $T_s=1451\pm263\rm\,K$. The dotted line indicates $\sigma_{\tau}(v)$. 
}
\vspace{-2pt}
\label{f:zoom}
\end{center}
\end{figure}

We model the remaining contribution to the expected profile by WNM not detected in absorption using a set of $K$ independent Gaussian functions. For each $k^{\mathrm{th}}$ component, we assume that a fraction $F_k$ lies in front of all $N$ absorption components, and thus a fraction $(1-F_k)$ is absorbed by the intervening gas. The emission brightness temperature from these $K$ non-absorbing Gaussian components is given by,

\begin{equation}
T_{B,\mathrm{EM}}(v)= \sum_{k=0}^{K-1} [F_{k} + (1-F_{k})e^{-\tau(v)}] \times T_{0,k}e^{-[(v-v_{0,k})/\delta v_{k}]^2 },
\label{tb:wnm}
\end{equation}

\noindent where ($T_{0,k},v_{0,k},\delta v_k$) are the (peak in units of brightness temperature, central velocity, 1/$e$ width) of the $k^{\mathrm{th}}$ emission component. For a given set of absorption components, the expected profile is fit for the spin temperatures of the $N$ absorption components and the Gaussian parameters of the $K$ emission components. This fit is repeated for all permutations of absorption components along the line of sight, for every value of $F_k$ for each emission component. Following the method of HT03, we allow $F_k$ to be 0, 0.5, or 1. The values of $F_k$ affect the derivation of spin temperatures, but any finer tuning of the values produces results that are difficult to distinguish statistically. Overall, there are $N!\times3^{K}$ permutations for each line of sight. The final fit selected is the one with the smallest residuals, and the final spin temperatures are calculated by a weighted average over all trials, where the weights are equal to the inverse of the variance of the residuals to the $T_{\rm exp}(v)$ fit (HT03). The error in the final spin temperature is estimated from the variation in $T_s$ with $F_k$, as described in Section 3.5 of HT03.

The expected emission profile (Equation~\ref{e:texp}) has been baseline-corrected, meaning that the contribution from diffuse radio continuum emission, including the cosmic microwave background (CMB) and Galactic synchrotron emission, has been removed. This value, $T_{\rm sky}$, is equal to $2.725\rm\,K$ from the CMB plus an estimate of the Galactic synchrotron background at the source position. Given that the spectral index of the synchrotron background is about 2.7, we divide the 408 MHz brightness temperature taken from the Haslam et al.\,(1982) Galactic survey by a factor of $(1420/408)^{2.7}$ to estimate the synchrotron contribution at 1420 MHz. The 21-SPONGE sources are mostly at high latitude, and therefore the values of $T_{\rm sky}$ are generally small, ranging between $2.76\rm\,K$ and $2.85\rm\,K$. To account for its presence, we add $T_{\rm sky}$ back to $T_{\rm exp}(v)$ before performing the fit. 

\subsection{Absorption Noise Spectra}
\label{s:noise}

The noise in H\textsc{i} absorption spectra is not constant as a function of velocity. H\textsc{i} emission at Galactic velocities will increase the system temperature of the VLA antennas, producing different noise properties across the absorption spectra. To estimate the ``noise spectra", $\sigma_{\tau}(v)$, of the VLA absorption profiles, we follow the methods discussed by Roy et al.\,(2013a). The spectral noise is composed of (1) on-source noise, $\sigma_{\mathrm{on}}(v)$ (varies with velocity) and (2) off-line noise from the frequency-switched bandpass solution, $\sigma_{\mathrm{BP}}$ (constant with velocity). We first estimate the RMS noise in off-line channels, $\sigma_{\tau}$ (see Table 2), which includes contributions from both $\sigma_{\mathrm{on}}(v)$ and $\sigma_{\mathrm{BP}}$. We then scale $\sigma_{\tau}$ by ($(T_{\mathrm{B}}(v) + T_{\mathrm{sys}})/T_{\mathrm{sys}}$) to calculate $\sigma_{\mathrm{on}}(v)$. The antenna used to produce the LAB survey is of similar size to an individual VLA antenna, so we use H\textsc{i} emission from the LAB survey to estimate the brightness temperature as a function of frequency for each source ($T_{\mathrm{B}}(v)$) and we assume that the system temperature of the VLA in the absence of H\textsc{i} emission is $T_{\rm sys}\sim25\rm\,K$. Finally, we calculate the noise spectrum by: $\sigma_{\tau}(v)^2=\sigma_{\mathrm{on}}(v)^2+\sigma_{\mathrm{BP}}^2$ (also see discussion by Roy et al.\,2013a). Examples of these noise spectra are displayed as dotted lines above and below the residuals of the Gaussian fits in Figures~\ref{3c123}, \ref{f:examples}, and \ref{f:zoom}. We use the noise spectra to visually inspect the fit residuals, not to quantify the goodness-of-fit.

\subsection{Component Selection: Absorption}

The process of Gaussian fitting has many inherent uncertainties, including the determination of number of fit components. The noise level in the fit residuals can be used as a goodness-of-fit indicator, but it cannot easily distinguish between different numbers of components, because adding more components will always reduce the residual noise level without necessarily bringing the fit any closer to an accurate depiction of reality.

To determine the best-fit number of Gaussian components to the absorption spectra, we begin by using a reduced chi-squared ($\chi^2_{\mathrm{red}}$) test. We calculate $\chi^2_{\mathrm{red}}$ using,

\begin{equation}
\chi^2_{\mathrm{red}} = \frac{1}{\nu}\sum\frac{(O(v)-M(v))^2}{\sigma_{\tau}^2},
\end{equation}

\noindent where $O(v)$ are the observed data, $M(v)$ are the fitted model data, $\sigma_{\tau}$ is off-line RMS noise in the absorption spectrum (from Table 2) and $\nu$ is the number of degrees of freedom (DOF: equal to number of measurements (256 spectral channels) minus the number of fitted parameters, which includes the height, width and center of each Gaussian function and a constant continuum offset).

For each profile, we would like to include the minimum number of components that brings $\chi^2_{\mathrm{red}}$ near to 1.0. However, in many cases, the difference in $\chi^2_{\mathrm{red}}$ between fits of different numbers of components is too small to confidently determine the best fit.  As an example, we present the case of 3C123 in Figure~\ref{3c123}. The central component of the absorption spectrum is very strong, and is poorly fit by a single Gaussian component. Including an additional broad component at the base of the central component does improve the fit, and including additional narrow lines improves the fit further. 

To compare the $\chi^2_{\mathrm{red}}$ values of different possible Gaussian fits, such as those shown for 3C123 in Figure~\ref{3c123}, we employ an F-test (e.g., Westmoquette et al.\,2007, Dawson et al.\,2011) using the IDL procedure mpftest.pro\footnote{http://purl.com/net/mpfit}. The F-test allows us to quantify the improvement in $\chi^2_{\mathrm{red}}$ as a function of added DOF to the Gaussian model. First, the procedure computes the F-value, which is defined as a ratio of $\chi^2_{\mathrm{red}}$ and $\nu$ values for two models:
\begin{equation}
\mathrm{F}=\frac{\chi^2_{\mathrm{red},1}/\nu_1}{\chi^2_{\mathrm{red,2}}/\nu_2},
\label{Fval}
\end{equation}

\noindent where the subscript ``1" denotes the base model which involves the fewest number of Gaussian components, and  ``2" denotes a model including an additional component (i.e., three additional DOF). The probability density function $p_f(f;\nu_1,\nu_2)$ for all F-values ($f$) described by DOF $\nu_1$ and $\nu_2$ is given by\footnote{NIST/SEMATECH e-Handbook of Statistical Methods, http://www.itl.nist.gov/div898/handbook/, 01/2015.},
\begin{equation}
p_f(f;\nu_1,\nu_2) = \frac{\Gamma[(\nu_1+\nu_2/2)]}{\Gamma(\nu_1/2)\Gamma(\nu_2/2)} \left ( \frac{\nu_1}{\nu_2} \right ) ^{\nu_1/2} \times \frac{f^{1/2(\nu_1-2)}}{(1+f\nu_1/\nu_2)^{1/2(\nu_1+\nu_2)}}.
\label{Fdistro}
\end{equation}

\noindent From this we calculate the probability that a value $f$ drawn from $p_f(f; \nu_1, \nu_2)$ equals or exceeds the given value of F (Equation~\ref{Fval}), which is given by,
\begin{equation}
P_\mathrm{F}(\mathrm{F};\nu_1,\nu_2)=\int_\mathrm{F}^\infty df~p_f(f;\nu_1,\nu_2), 
\end{equation}

\noindent where $P_F$ (or confidence level $C_F=1-P_F$) is a measure of the probability that the added component merits inclusion in the model. 

For the majority of sources, the inclusion of additional Gaussian components results in $C_F>0.99$ before $C_F$ drops far below $C_F\sim0.9$ and indicates low confidence in the most recent component-addition. In three cases (sources 3C111, 3C133 and 3C459), the fit parameters varied strongly with the initial guesses, and we observed that the fits remained stable between the final component-additions if we applied a cutoff value of $C_F=0.97$. Therefore, to apply a uniform standard to all sources, we use a strict threshold of $C_F=0.97$ for including components, noting that in all but three cases, the effective cutoff is $C_F>0.99$. 

As an example, in the case of 3C123, the confidence level of including one additional component (Figure~\ref{3c123}b) to the base fit (Figure~\ref{3c123}a) is $C_F=0.99$, indicating that the additional component exists with high confidence. The confidence level of including a second additional component (Figure~\ref{3c123}c) in addition to the new fit (Figure~\ref{3c123}b) is $C_F=0.61$, and therefore we reject this additional component.

\subsection{Component Selection: Emission}

The selection of Gaussian components for the expected emission profiles is more difficult.  Taking the absorption line fits as fixed in central velocity, peak optical depth and line width (FWHM), we simultaneously solve for the Gaussian parameters of additional emission components and the spin temperatures of the absorption components given emission information along the line of sight (Equations~\ref{tb:cnm} and~\ref{tb:wnm}). Considering the effects of stray radiation on the emission profiles and the complexity of the fits, we select a minimum number of components that brings the fit residuals below the $3\sigma_{T_{\mathrm{exp}}}(v)$ level. We then apply a similar F-test as in the absorption component selection and add additional components until the confidence level of the fit improvement drops below the cutoff ($C_F=0.97$). 

Given the final selection of Gaussian components, we then solve the radiative transfer equations again for all possible orderings of the $N$ absorption components along the line of sight. We repeat this line of sight iteration for all permutations of $F_k$, the fraction of WNM unabsorbed by intervening CNM (see Section 3.1) among the $K$ emission components. If the total number of absorption components was large (defined here as greater than or equal to 6), we only varied the order of the most blended components at the center of the spectrum. The effect of different CNM cloud order on $T_s$ is very small compared with the effect of different values of $F_k$ per emission component, therefore we held the more isolated components in these complex spectra fixed in their LOS order with little consequence to the derived $T_s$ values for the overall fit.

\section{Selected Examples}
\label{s:examples}

To illustrate the fitting method further, we select four example sources and display their spectra and Gaussian decompositions in Figure~\ref{f:examples}. The top panel of each plot contains the emission profile, $T_{\mathrm{exp}}(v)$, from HT03, a2770, or LAB (thin solid), the total fit to $T_{\mathrm{exp}}(v)$ (thick solid), the total absorption contribution to the emission profile, $T_{\rm B,\,ABS}(v)$ (Equation 1, dashed), and the total emission-only contribution, $T_{\rm B,\,EM}(v)$ (Equation 1, triple dot-dash). The bottom panel of each plot contains the VLA absorption profile (thin solid), total fit to the absorption profile (thick solid), and individual Gaussian components (dashed). The residuals of each fit are plotted at the bottom of each panel, with noise envelopes $\sigma_{T_{\rm exp}}(v)$ (Section 2.4) and $\pm\sigma_{\tau}(v)$ (Section 3.2) to illustrate the goodness of fit (dotted). 

We compare in detail our Gaussian decomposition with overlapping H\textsc{i} absorption line surveys, including HT03, Mohan et al.\,(2004)  and \citet{Roy13b}. All three surveys fit Gaussian functions to their absorption lines and followed different strategies. For example, \citet{Roy13b} fit increasing numbers of Gaussian functions until $\chi_{\rm red}^2$ for the fit was as close to 1 as possible. This resulted in a minimum of 1, maximum of 20 and of median of 7 Gaussian components over their 30 spectra. This approach requires detailed knowledge of the spectral noise properties, as well as the assumption that Gaussian functions are the proper representation of all components, which can fail if line broadening is not due to thermal and/or turbulence motions. Our method also makes these assumptions, but we fit fewer components per spectrum (minimum of 1, maximum of 11 and median of 5) to not ``over-fit" at the expense of inferior $\chi_{\rm red}^2$ values for the fits. In another approach, HT03 stress that human judgment is necessary for producing reliable Gaussian fits. For example, two fitted components with similar central velocities and FWHMs could in reality be a single component with a non-Gaussian intrinsic lineshape. No Gaussian decomposition is a unique solution, and it is impossible to prove which solution is most physically relevant, but the differences between our respective Gaussian fitting strategies will affect the comparison of our results. We emphasize that our F-test method for component selection imposes a uniform standard to a non-unique fitting problem, and aims to make our decisions reproducible.

\subsection{3C138 (Figure~\ref{f:examples}a)}

3C138 is included in the Millennium Arecibo survey (HT03), and they identified many similar, strong components, as we do, in their Gaussian decomposition. As seen in Figure~\ref{f:examples}a, in addition to narrow cold components at velocities 9.1, 6.4, 1.6 and -0.5 km/s, we identify a shallow broad underlying component, centered at 1.9 $\pm$ 0.3 km/s with a FWHM of $15.1\pm0.6\rm\,km\,s^{-1}$. This corresponds to a maximum kinetic temperature of $T_{k,\rm max}=5000\rm\,K$, and with a fit to the emission profile (from HT03), corresponds to a spin temperature of $456\pm40\rm\,K$. The maximum spin temperature derived by HT03 for an absorption component in the direction of this source is $T_s=380\pm23\rm\,K$, corresponding to a similar component at $1.8\rm\,km\,s^{-1}$. 

This source is also included in the Roy et al.\,(2013b) WSRT/GMRT/ATCA survey. Whereas we fit a total of 6 components before the fit ceases to improve significantly by the F-test, Roy et al.\,(2013b) fit 13 components. Given the complex nature of the spectrum, and the large difference in total numbers of components, there are few closely matching components between the two decompositions.

\subsection{3C286 (Figure~\ref{f:examples}b)}

With a sensitivity of $\sigma_{\tau}=9\times10^{-3}$, HT03 were unable to detect H\textsc{i} absorption in the direction of 3C286. Stanimirovi\'c \& Heiles (2005) demonstrate that with additional integration times at the Arecibo Observatory, weak absorption lines below the HT03 sensitivity threshold are easily detected, which can have important implications for HT03's measured fractions of H\textsc{i} in the CNM, WNM and thermally unstable phases. For 3C286, they find 3 velocity components at $-28.8$, $-14.3$ and $-7.4\rm\,km\,s^{-1}$, with $T_{k,\rm max}=106$, $115$ and $315\rm\,K$ and $T_s=89\pm7$, $37\pm4$ and $30\pm20\rm\,K$. We find similar velocity components, at $-28.5$, $-14.3$ and $-7.3\rm\,km\,s^{-1}$, with $T_{k,\rm max}=116$, $220$ and $405\rm\,K$ and $T_s=63\pm13$, $84\pm12$ and $27\pm9\rm\,K$. Roy et al.\,(2013b) find the same three components, in addition to a broad line centered at $-8.4\rm\,km\,s^{-1}$ with peak optical depth $\tau=0.001$ and $T_{k,\rm max}=5140\rm\,K$. Their achieved sensitivity of $\sigma_{\tau}=3\times10^{-4}$ allows them to detect this weak component whereas our sensitivity of $\sigma_{\tau}=7\times10^{-4}$ puts it below our detection limit. We note that they do not discuss the effect of bandpass instability on their detection of broad, weak lines. 

\subsection{4C16.09 (Figure~\ref{f:examples}c)}

Although 4C16.09 is not included in the Millennium survey (HT03), it is included in the Mohan et al.\,(2004) GMRT survey of H\textsc{i} absorption. They achieve a mean survey sensitivity in optical depth of $\sigma_{\tau} \sim 3\times10^{-3}$ and the spectra have velocity resolution of $2.1\rm\,km\,s^{-1}$ over roughly $350\rm\,km\,s^{-1}$. Our absorption profile agrees well in shape with their results. They detect two narrow absorption components, at velocities $7.5$ and $0.2\rm\,km\,s^{-1}$ with corresponding spin temperatures $T_s=30\rm\,K$ and $45\rm\,K$ respectively (Mohan et al.\,2004). We detect components at similar velocities, $7.6$ and $0.7\rm\,km\,s^{-1}$, in addition to four additional absorption components (see Figure~\ref{f:examples}c). We identify two additional, broad emission components which are prominent in the expected emission profile and produce a maximum spin temperature of $T_s=240\pm10\rm\,K$. In comparison, Roy et al.\,(2013b) have the same sensitivity of $\sigma_{\tau}=8\times10^{-4}$ and find 8 Gaussian components. 

\subsection{PKS0531+19 (Figure~\ref{f:examples}d)}

The profile in the direction of PKS0531+19 is highly complex. Detected by HT03 in the Millennium Survey, they identify four narrow absorption components at velocities $9.6$, $5.5$, $1.8$ and $-23.1\rm\, km\,s^{-1}$ (HT03). We detect features at similar velocities, in addition to two broad, shallow components centered at $21$ and $-25\rm\, km\,s^{-1}$ corresponding to spin temperatures of $T_s=1129\pm74\rm\,K$ and $1451\pm263\rm\,K$. In Figure~\ref{f:pks_zoom} we zoom in on the second of these components, shown by the thick green line. This source is a good example of the benefit of high sensitivity. The RMS noise in the absorption profile is $\sigma_{\tau}=5\times10^{-4}$, so that the broad component with peak optical depth equal to $\tau=0.0015$ is a $3\sigma$ detection. HT03 achieve a sensitivity of $\sigma_{\tau}\sim0.0015$ in their absorption profile towards this source, and therefore did not fit a Gaussian function to this line given that it fell below their $1\sigma$ level. 

\section{Properties of cold and warm neutral gas}

Following complete decomposition of all absorption and emission profiles with Gaussian functions, we can use estimates of temperature and column density to infer properties of the CNM and WNM. 

\subsection{Temperature}

\begin{figure}
\begin{center}
\vspace{-15pt}
\includegraphics[width=0.5\textwidth]{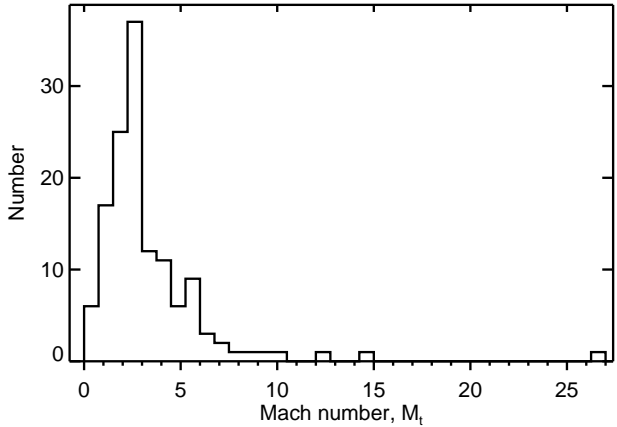}
\vspace{-140pt}
\caption{\label{f:mach}
Histogram of turbulent Mach number $M_t$ for all absorption-detected Gaussian components, calculated from Equation~(\ref{e:mach}) (Heiles \& Troland 2003b Equation 17).}
\label{f:tempcompare}
\end{center}
\end{figure}

The FWHM ($\Delta v$) of each Gaussian function determines its maximum kinetic temperature, $T_{k,\rm max}$, by

\begin{equation}
T_{k,\rm max}=m_H/(8 k_B\,\mathrm{ln}(2) )\times \Delta v^2=21.866\times\Delta v^2,
\label{e:tk}
\end{equation}

\noindent for hydrogen mass $m_H$, Boltzmann's constant $k_B$, and FWHM $\Delta v$ in $\rm km\,s^{-1}$ (e.g., HT03). 
This quantity contains contributions from both thermal broadening and turbulent gas motions so that,
\begin{equation}
T_{k,\rm max}= T_k + \frac{m_H v_{\rm turb}^2}{2 k_B},
\label{e:tkmax}
\end{equation}

\noindent where $T_k$ is the true kinetic temperature of the gas (e.g. Liszt 2001). Therefore, $T_{k,\rm max}$ is an upper limit to the kinetic temperature in the presence of turbulent broadening. As shown by Heiles \& Troland (2003b), nonthermal motions in the CNM are characterized by a turbulent Mach number ($M_t$) of $M_t\sim3$, implying that CNM gas is supersonic. To estimate $M_t$, we solve Equation~(\ref{e:tkmax}) for $v_{\rm turb}^2$, and then assume that this line of sight turbulent velocity is $\sqrt(3)$ times the three-dimensional turbulent velocity ($v_{\rm turb, 3D}$). The turbulent Mach number $M_t$ is given by $v_{\rm turb, 3D}^2/C_s^2$, where $C_s$ is the isothermal sound speed. We use $C_s$ because thermal equilibrium is established quickly in the CNM, which dominates the absorption profiles (Heiles \& Troland 2003b). Following Heiles \& Troland (2003b), we adopt $C_s^2=kT_s/1.4m_{\rm H}$ for mean atomic weight $1.4m_{\rm H}$. Therefore, the turbulent Mach number is given by, 
\begin{equation}
M_t^2=\frac{v_{\rm turb,3D}^2}{C_s^2}=4.2\left(\frac{T_{k,\rm max}}{T_s}-1\right),
\label{e:mach}
\end{equation}

\noindent which is Equation (17) from Heiles \& Troland (2003b). In Figure~\ref{f:mach}, we display a histogram of $M_t$ for all 21-SPONGE absorption-detected Gaussian components, which has a peak at $M_t\sim3$ (median $M_t=2.86$) and agrees with the results from HT03. The distribution is strongly peaked, and $40\%$ of all components have Mach numbers between 2 and 4. 

The spin temperature is determined by the ambient radiation field, collisions between hydrogen atoms, electrons and protons, and Ly$\alpha$ scattering. For the CNM, high densities allow the collisional contribution to dominate 21-cm excitation so that $T_s$ is expected to be equal to $T_k$. At low densities, i.e. in more diffuse WNM gas, collisions are not sufficient to thermalize the 21-cm transition and therefore $T_s$ is expected to be less than $T_k$ (e.g. Liszt 2001).

In the case of 12 sources, one or more of the Gaussian functions fit to the absorption spectrum did not have a corresponding spectral feature at the same velocity in emission (likely due to the cold component occupying a very small solid angle, Heiles \& Troland 2003a). Therefore, the converged $T_s$ values for these components are unreasonably low, $<1\rm\,K$. Given the uncertain presence of stray radiation in the Arecibo emission profiles, we take a conservative approach for estimating $T_s$ for these components. We adopt the median $M_t$ value measured over all fitted Gaussian functions, $M_t=2.86$ and solve Equation~(\ref{e:mach}) for $T_s$ and its error given $T_{k,\rm\,max}$ and its error. 

\begin{figure*}
\begin{center}
\vspace{-5pt}
\includegraphics[width=\textwidth]{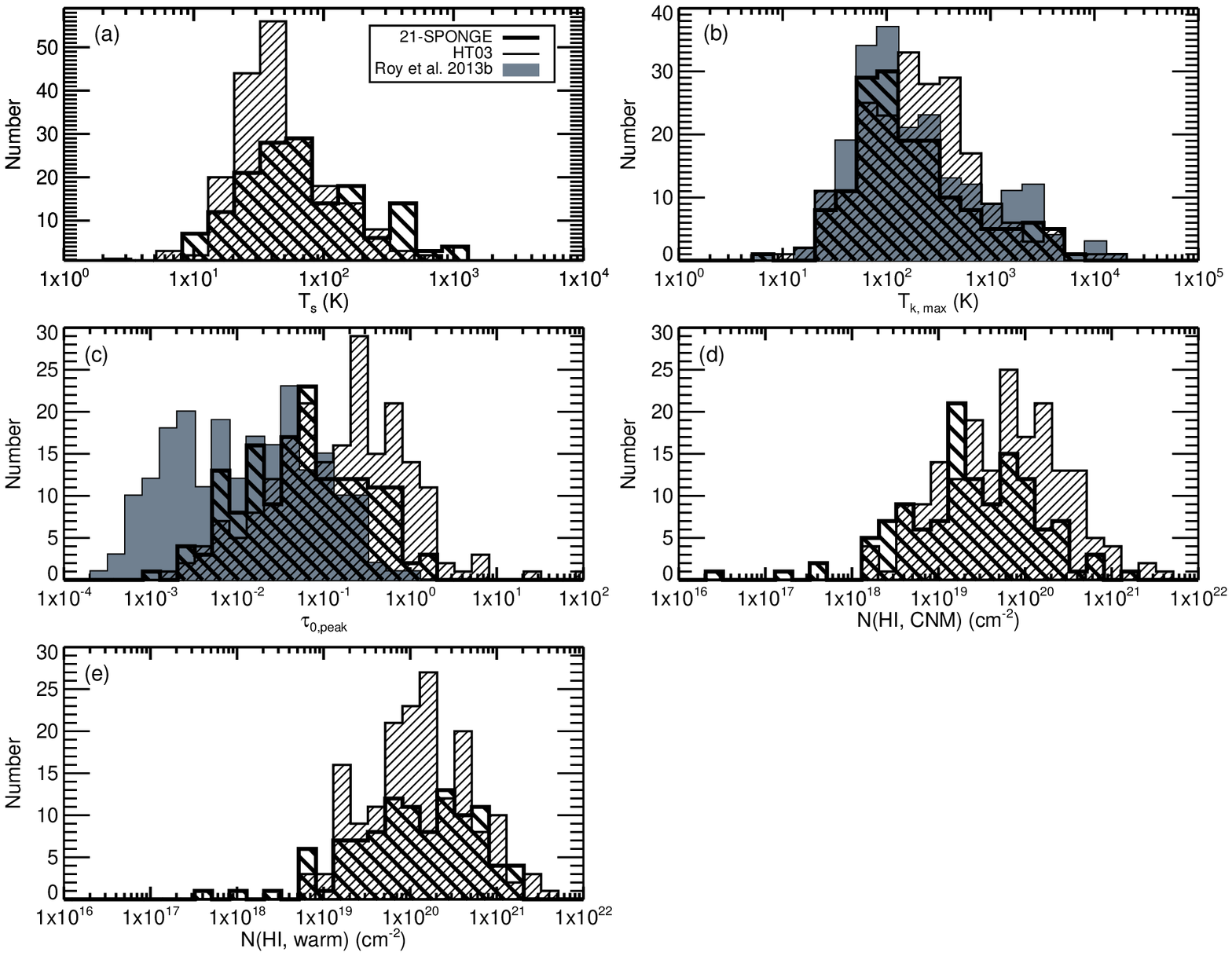}
\vspace{-240pt}
\caption{\label{f:hist}
Histograms of various properties from the Gaussian fits to the current \NoOfSources{} sources in 21-SPONGE (thick black), including: (a) $T_s\rm\,(K)$, (b) $T_{\rm k,max}\rm\,(K)$, (c) peak optical depth in absorption, $\tau_{0,\rm\,peak}$, (d) $\rm N(\rm H\textsc{i}, CNM)$: column density of all CNM components, defined as those with $T_s<200\rm\,K$, (d) $\rm N(\rm H\textsc{i}, warm)$: column density of all ``warm" components, defined as those with $T_s>200\rm\,K$ and those detected only in emission. The results for the Millennium Survey (HT03, thin black) and the WSRT/GMRT/ATCA survey by Roy et al.\,(2013b) (shaded gray) are included for comparison. Results from Roy et al.\,(2013b) are not included in panels a, d and e because they do not compute these parameters for individual Gaussian components.
}
\vspace{20pt}
\end{center}
\end{figure*}

\subsection{Column Density}

We estimate the H\textsc{i} column density of the absorbing H\textsc{i} using the expression:
\begin{equation}
\mathrm{N}(\mathrm{H\textsc{i}, ABS}) = C_0~T_s \int \tau(v)~\mathrm{d}v
\label{e:nhcnmtot}
\end{equation}

\noindent where $C_0=1.823 \times 10^{18}\,\mathrm{cm^{-2}\,K^{-1}\,(km\,s^{-1})^{-1}}$, $T_{\rm s}$ is the derived spin temperature and $\tau(v)$ is the optical depth as a function of velocity in km s$^{-1}$. For each $n^{\rm th}$ absorption component, we use parameters from the Gaussian fit to calculate the column density, so that,
\begin{equation}
\mathrm{N}(\mathrm{H \textsc{i}, ABS})_n=1.064 \times C_0 \times T_{s,n}  \times \tau_{0,\mathrm{peak},n} \times \Delta v_n,
\label{e:nhcnm}
\end{equation}

\noindent where ($T_{s,n}, \tau_{0,\mathrm{peak},n}, \Delta v_n$) are the (spin temperature, peak optical depth and FWHM in km s$^{-1}$) of the $n^{\rm th}$ component and 1.064 converts the product to the area under a Gaussian function with the given height and width.
For the emission components not detected in absorption, we estimate the column density as,
\begin{equation}
\mathrm{N}(\mathrm{H\textsc{i}, EM}) = C_0 \int T_{\mathrm B}(v)~\mathrm{d} v,
\end{equation}

\noindent where $T_{\rm B}$ is the brightness temperature. Using the Gaussian fit to the emission profile, we estimate the column density associated with each $k^{\rm th}$ non-absorbing component as,

\begin{equation}
\mathrm{N}(\mathrm{H\textsc{i}, EM})_{\rm k}=1.064 \times C_0\times T_{0,k} \times \Delta v_k,
\label{e:nhwnm}
\end{equation}

\noindent where $T_{0,k}$ and $\Delta v_k$ are the peak and FWHM in km s$^{-1}$ of the $k^{\rm th}$ component.

\begin{figure}
\begin{center}
\includegraphics[width=0.5\textwidth]{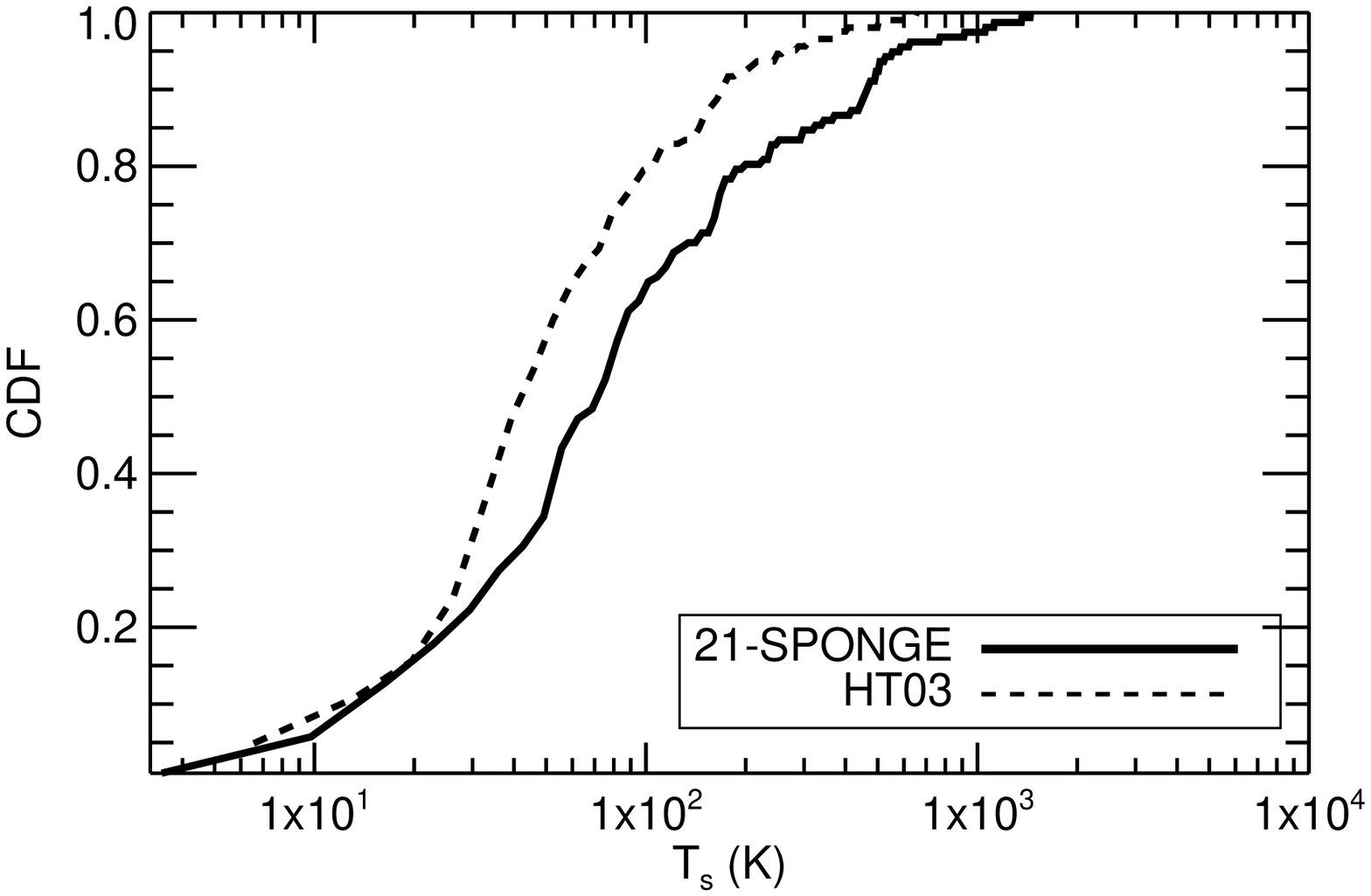}
\vspace{-120pt}
\caption{
Cumulative distribution function of the $T_s$ values shown in Figure~\ref{f:hist}a for 21-SPONGE and HT03. }
\label{f:ts_cg}
\end{center}
\end{figure}

\begin{figure}
\begin{center}
\includegraphics[width=0.5\textwidth]{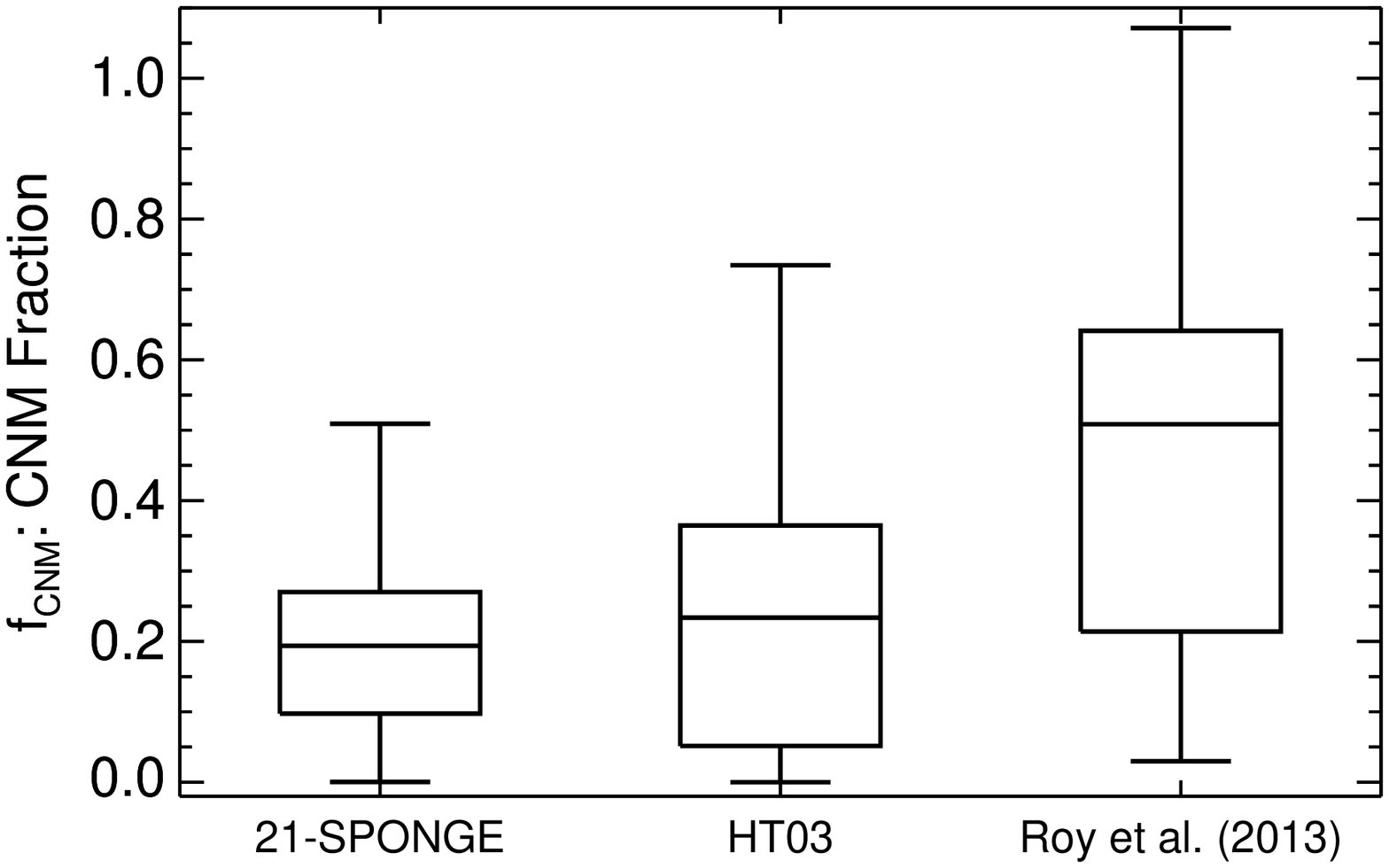}
\vspace{-120pt}
\caption{
``Box and whisker" plot comparing the CNM fractions ($f_{\rm CNM}$) for 21-SPONGE with HT03 and Roy et al.\,(2013). Each box spans the $25^{\rm th}$ through $75^{\rm th}$ percentiles of each distirbution, the median value is indicated by the horizontal line within each box, and the bars extend to the minimum and maximum values.
}
\label{f:nhlos}
\end{center}
\end{figure}

\section{Comparison with previous surveys}

In Figure~\ref{f:hist}, for all absorption-detected Gaussian functions we display histograms of: (a) spin temperature ($T_s$), (b) maximum kinetic temperature ($T_{k,\rm max}$), (c) peak optical depth ($\tau_{\rm 0,peak}$), (d) column density for all CNM clouds, defined as those with $T_s\leq 200\rm\,K$, ($\rm N(\rm H\textsc{i}, CNM)$), and finally (e) column density for all warm clouds, defined as those emission-only components and absorption-detected components with $T_s\geq200\rm\,K$ ($\rm N(\rm H\textsc{i}, warm)$). We overlay the same quantities for the 79 sources from the Millennium survey (HT03) in the panels of Figure~\ref{f:hist} for comparison. 

{
\setlength{\extrarowheight}{1.2pt}
\setlength{\tabcolsep}{12pt}
\begin{table*}
\caption{Line of sight column density parameters}
\centering
\label{tab:los}
\begin{tabular}{lrrrc}
\hline
\hline
\small
Source  & $\Sigma \rm N(\rm H\textsc{i}, CNM)_{20}$   &  $\Sigma \rm N(\rm H\textsc{i}, warm)_{20}$ & $\rm N(\rm H\textsc{i}, total)_{20}$  & $f_{\rm CNM}$ \\
        &  ($10^{20}\rm\,cm^{-2}$)           &  ($10^{20}\rm\,cm^{-2}$)            &  ($10^{20}\rm\,cm^{-2}$)      &          \\
\hline
4C32.44&0.17 $\pm$ 0.09&1.19 $\pm$ 0.03&1.36 $\pm$ 0.09&0.12 $\pm$ 0.06 \\ 
3C286&0.06 $\pm$ 0.02&1.28 $\pm$ 0.09&1.34 $\pm$ 0.09&0.05 $\pm$ 0.02 \\ 
4C12.50&0.36 $\pm$ 0.11&2.01 $\pm$ 0.03&2.37 $\pm$ 0.11&0.15 $\pm$ 0.05 \\ 
3C273&0.05 $\pm$ 0.19&2.32 $\pm$ 0.03&2.37 $\pm$ 0.19&0.02 $\pm$ 0.08 \\ 
3C298&0.06 $\pm$ 0.37&2.33 $\pm$ 0.09&2.39 $\pm$ 0.38&0.03 $\pm$ 0.15 \\ 
4C04.51&0.24 $\pm$ 0.06&4.31 $\pm$ 0.11&4.55 $\pm$ 0.13&0.05 $\pm$ 0.01 \\ 
3C237&0.17 $\pm$ 0.20&2.43 $\pm$ 0.11&2.60 $\pm$ 0.23&0.07 $\pm$ 0.08 \\ 
3C225A&1.19 $\pm$ 0.30&3.05 $\pm$ 0.06&4.24 $\pm$ 0.30&0.28 $\pm$ 0.07 \\ 
3C225B&0.98 $\pm$ 0.18&3.57 $\pm$ 0.11&4.55 $\pm$ 0.21&0.22 $\pm$ 0.04 \\ 
3C345$^a$&0.00 $\pm$ 0.00&1.01 $\pm$ 0.13&1.01 $\pm$ 0.13&0.00 $\pm$ 0.00 \\ 
3C327.1&1.76 $\pm$ 1.17&6.96 $\pm$ 0.11&8.72 $\pm$ 1.18&0.20 $\pm$ 0.14 \\ 
3C147&7.44 $\pm$ 1.79&13.15 $\pm$ 0.78&20.59 $\pm$ 1.95&0.36 $\pm$ 0.09 \\ 
4C33.48&1.54 $\pm$ 3.00&14.22 $\pm$ 0.48&15.76 $\pm$ 3.04&0.10 $\pm$ 0.19 \\ 
3C154&9.75 $\pm$ 14.94&36.19 $\pm$ 0.49&45.94 $\pm$ 14.94&0.21 $\pm$ 0.33 \\ 
3C410&30.39 $\pm$ 6.64&29.29 $\pm$ 0.81&59.69 $\pm$ 6.69&0.51 $\pm$ 0.13 \\ 
B2050&3.00 $\pm$ 2.02&30.86 $\pm$ 1.33&33.86 $\pm$ 2.42&0.09 $\pm$ 0.06 \\ 
3C409&5.33 $\pm$ 3.49&26.36 $\pm$ 0.55&31.69 $\pm$ 3.53&0.17 $\pm$ 0.11 \\ 
PKS0531&8.39 $\pm$ 9.61&20.45 $\pm$ 1.48&28.84 $\pm$ 9.72&0.29 $\pm$ 0.35 \\ 
3C111&3.99 $\pm$ 3.78&32.44 $\pm$ 2.29&36.43 $\pm$ 4.42&0.11 $\pm$ 0.10 \\ 
3C133&9.76 $\pm$ 4.02&26.22 $\pm$ 0.46&35.98 $\pm$ 4.05&0.27 $\pm$ 0.12 \\ 
3C138&6.01 $\pm$ 2.83&20.42 $\pm$ 0.58&26.43 $\pm$ 2.89&0.23 $\pm$ 0.11 \\ 
3C123&5.15 $\pm$ 4.41&19.41 $\pm$ 1.90&24.55 $\pm$ 4.80&0.21 $\pm$ 0.18 \\ 
3C433&1.15 $\pm$ 2.09&9.23 $\pm$ 0.13&10.38 $\pm$ 2.10&0.11 $\pm$ 0.20 \\ 
3C120&10.86 $\pm$ 2.24&11.36 $\pm$ 0.26&22.22 $\pm$ 2.26&0.49 $\pm$ 0.11 \\ 
3C48&0.70 $\pm$ 0.36&4.25 $\pm$ 0.19&4.95 $\pm$ 0.41&0.14 $\pm$ 0.07 \\ 
4C16.09&3.27 $\pm$ 1.10&8.83 $\pm$ 0.24&12.10 $\pm$ 1.13&0.27 $\pm$ 0.09 \\ 
3C454.3&1.99 $\pm$ 15.74&5.92 $\pm$ 0.63&7.90 $\pm$ 15.75&0.25 $\pm$ 2.05 \\ 
J2232&1.21 $\pm$ 0.37&4.29 $\pm$ 0.17&5.50 $\pm$ 0.41&0.22 $\pm$ 0.07 \\ 
3C78&3.81 $\pm$ 1.91&9.93 $\pm$ 0.27&13.74 $\pm$ 1.93&0.28 $\pm$ 0.14 \\ 
3C459&1.17 $\pm$ 0.49&5.12 $\pm$ 0.04&6.29 $\pm$ 0.49&0.19 $\pm$ 0.08 \\ 
\hline
\end{tabular}
\vskip 0.1 in
\footnotesize
\raggedright
$^a$: $N(\rm H\textsc{i}, CNM)_{20}=3\pm1\times10^{16}\,cm^{-2}$
\end{table*}
}
The superior sensitivity of 21-SPONGE allows us to detect absorption lines with smaller $\tau_{\rm peak}$ and/or higher $T_s$ than in HT03, which explains the poor agreement between the 21-SPONGE and HT03 distributions in Figures~\ref{f:hist}a and \ref{f:hist}c. We are therefore more sensitive to low-$\rm N(\rm H\textsc{i}, CNM)$, as shown in Figure~\ref{f:hist}d and in Equation~(\ref{e:nhcnm}). 

In Figure~\ref{f:hist}a, we detect a strong tail of high-$T_s$ components in comparison with HT03. This results in our median $T_s=77\rm\,K$, which is higher than their median value of $T_s=48\rm\,K$. The maximum $T_s$ we detect is $T_s=1451\pm263\rm\,K$, compared with the maximum value measured by HT03 of $T_s=656\rm\,K$, demonstrating that it is possible to directly measure $T_s$ from warm gas with improved sensitivity in H\textsc{i} absorption. 

However, although we have the sensitivity to detect H\textsc{i} at higher spin temperatures than $1500\rm\,K$ directly in absorption, we only detect 32/\NoAbsComps{} Gaussian components above the expected range for the CNM ($T_s=40-200\rm\,K$ from Wolfire et al.\,2003), and only 4/\NoAbsComps{} above $T_s=1000\rm\,K$. This is a surprising result from 21-SPONGE so far, given that HT03 estimate (based on indirect measurements from H\textsc{i} emission only) that $48\%$ of components should be in the thermally unstable range of $500\leq T_s\leq 5000\rm\,K$. Our current estimate for the thermally unstable fraction in this same range-- $\sim20$\% by number-- is more consistent with estimates from colliding flow models (e.g., Audit \& Hennebelle 2005).

In Figure~\ref{f:ts_cg}, we present cumulative distribution functions of the data in Figure~\ref{f:hist}a. A K-S test indicates that the two distributions are highly unlikely to be drawn from the same parent population $(\mathrm{K}$-$\mathrm{S}=0.23, \mathrm{p}=0.00$). The 21-SPONGE distibrution is much shallower at high temperature, indicating again that we are detecting more gas at higher $T_s$ than was done by HT03 using similar analysis methods.

In Figures~\ref{f:hist}b and \ref{f:hist}c we also include Gaussian decomposition results from the WSRT/GMRT/ATCA survey presented in Roy et al.\,(2013b) (grey shaded histograms; see Table 1 for survey information). Our two surveys have similar sensitivity in absorption. Following Gaussian decomposition of their absorption spectra, Roy et al.\,(2013b) do not estimate $T_s$, $N(\rm H\textsc{i}, CNM)$ or $N(\rm H\textsc{i}, warm)$ for their Gaussian components. As discussed in Section~\ref{s:examples}, Roy et al.\,(2013b) fit many more components per spectrum than we do, and therefore they detect many more low-$\tau$ lines than either we or HT03 do, as seen by the strong low-$\tau$ tail in the gray-shaded histogram in Figure~\ref{f:hist}c. 

We agree well with both Roy et al.\,(2013b) and HT03 in $T_{k,\rm max}$, shown in Figure~\ref{f:hist}b. From Equation~(\ref{e:tk}), this implies that we are sensitive to similar absorption line widths. We find a median value of $T_{k,\rm max}=160\rm\,K$, compared with HT03's median value of $T_{k,\rm max}=220\rm\,K$ and Roy et al.\,(2013b)'s value of $T_{k,\rm max}=160\rm\,K$. We find that $5.7\%$ of components have $T_{k,\rm max}\leq 40\rm\,K$, which is below the minimum theoretically expected kinetic temperature for the CNM (e.g. Wolfire et al.\,2003), compared wth 4.7\% by HT03 and 7.0\% by Roy et al.\,(2013). At the extreme end of the low-$T_{k,\rm max}$ values, we find one weak component with $T_{k,\rm max}=9\pm4\rm\,K$ in the direction of 3C345. These low-temperature components can be explained by ineffective photoelectric heating from dust and polycyclic aromatic hydrocarbons (PAHs) along those sightlines, which allows for the existence of very cold clouds with $T_{k, \rm max}\sim10\rm\,K$ (Wolfire et al.\,1995, Heiles \& Troland 2003b, Roy et al.\,2013b). 

In Figures~\ref{f:hist}d and \ref{f:hist}e, it is clear that we generally agree with HT03 in measurements of $\rm N(\rm H\textsc{i}, CNM)$ and $\rm N(\rm H\textsc{i}, warm)$. However, our superior sensitivity in optical depth allows us to detect lines with much lower $\rm N(\rm H\textsc{i}, CNM)$. For example, we are able to detect an individual absorption line with $\rm N(\rm H\textsc{i}, CNM)=3\pm1\times10^{16}\rm\,cm^{-2}$, whereas HT03 measure a minimum column density of $2\times10^{18}\rm\,cm^{-2}$ for individual CNM clouds. An important result from the final 21-SPONGE data release will be constraints on the low-$\rm N(H\textsc{i})$ end of the CNM column density distribution. McKee \& Ostriker (1977) predicted a lower-limit to CNM column density of $\rm N(\rm H\textsc{i}, CNM)\sim10^{17}-10^{18}\rm\,cm^{-2}$, derived from balancing radiative losses and conductive heating at the interface between the CNM and WNM at various possible temperatures. As discussed by Stanimirovi\'c \& Heiles (2005), observational constraints on this lower limit (and its prevalence) can distinguish between different formation and evolution histories of these structures, whether by evaporation into the surrounding medium, or transient, dynamic formation at shock interfaces, or condensation from turbulent flows. 

\subsection{Line of sight properties}
\label{s:los}

In Table~\ref{tab:los} we compute line of sight column density-related quantities for the current 21-SPONGE sources, where the subscript ``20" indicates units of $10^{20}\rm\,cm^{-2}$.
These quantities include: (a) the total CNM column density ($\Sigma \rm N(H\textsc{i}, CNM)_{20}$), equal to the sum of the column densities of the absorption-detected Gaussian components with $T_s<200\rm\,K$; (b) the total WNM column density ($\Sigma \rm N(H\textsc{i}, warm)_{20}$), equal to the sum of the column densities of the absorption-detected Gaussian components with $T_s\geq200\rm\,K$ plus the column densities of all emission-only Gaussian components (which is a combination of both thermally-stable WNM and thermally-unstable gas); (c) the total H\textsc{i} column density per line of sight ($\rm N(H\textsc{i},total)_{20}=\Sigma \rm N(H\textsc{i}, CNM)_{20}+\Sigma \rm N(H\textsc{i}, warm)_{20}$); (d) the CNM fraction, $f_{\rm CNM}$, defined as:

\begin{equation}
f_{\rm CNM}=\frac{\Sigma \rm N(H\textsc{i}, CNM)}{\rm N(H\textsc{i},total)}.
\label{e:fcnm}
\end{equation}

To compare our results, we obtained or calculated the same quanitites for the HT03 and Roy et al.\,(2013b) samples.
For HT03, we used their published fit information to calculate all quantities in the same manner as for 21-SPONGE, described above. To calculate $\Sigma \rm N(H\textsc{i}, CNM)_{20}$, \citet{Roy13b} began by assuming that all absorbing gas is from the CNM with $T_s=200\rm\,K$ and solved Equation~(\ref{e:nhcnmtot}) to produce upper limits to $\Sigma \rm N(H\textsc{i}, CNM)_{20}$ for each line of sight. We estimated $\Sigma \rm N(H\textsc{i}, warm)_{20}$ from their results by subtracting $\Sigma \rm N(H\textsc{i}, CNM)_{20}$ from $N(\rm H\textsc{i}, total)_{20}$, which they calculated using the ``isothermal" method (Spitzer 1978, Dickey \& Benson 1982, Chengalur et al.\,2013).

We generally agree well with the results from HT03 and \citet{Roy13b}. All three studies measure the same range in $\Sigma \rm N(H\textsc{i}, CNM)_{20}$, between $\sim0.01$ and $\sim10$. We find a median $\Sigma \rm N(H\textsc{i}, CNM)_{20}=1.76$, compared with $1.45$ by HT03 and $1.76$ by Roy et al.\,(2013). From Figure~\ref{f:ts_cg}, we detect many more components in absorption with $T_s>200\rm\,K$, which have high $N(\rm H\textsc{i},ABS)_n$ according to Equation~(\ref{e:nhcnm}) and which will increase the total $\Sigma \rm N(H\textsc{i}, warm)_{20}$ along the line of sight. We measure median $\Sigma \rm N(H\textsc{i}, warm)_{20}=8.83$, which is much higher than HT03's median value of $3.97$ and Roy et al.\,(2013)'s median value of $1.41$. We measure a median total column density per LOS of $N(\rm H\textsc{i},total)_{20}=10.4$ compared with $5.54$ by HT03 and $4.33$ by Roy et al.\,(2013). 

Figure~\ref{f:nhlos} displays a ``box and whisker" plots of $f_{\rm CNM}$ for 21-SPONGE, HT03 and Roy et al.\,(2013a). By inspection, our results are consistent with the $f_{\rm CNM}$ distribution found by HT03 and Roy et al.\,(2013a). HT03 do not detect H\textsc{i} absorption in the direction of 15/79 sources, resulting in many values of $f_{\rm CNM}=0$. Whereas we only have one non-detection, we find 9/31 sources with $f_{\rm CNM}\leq0.1$. We measure median $f_{\rm CNM}=0.20$ compared with $0.23$ by HT03. We appear to find lower $f_{\rm CNM}$ than \citet{Roy13b}. Out of our \NoOfSources{} sources, we find a maximum $f_{\rm CNM}\sim0.51$, whereas \citet{Roy13b} find much higher values, all the way up to $f_{\rm CNM}\sim1$ with median $0.51$. As discussed above, their $\Sigma \rm N(H\textsc{i}, CNM)_{20}$ are strong upper limits, relying on the assumption that all absorbing gas has $T_s=200\rm\,K$, which neglects gas along the line of sight at higher and lower temperatures. Therefore, their $f_{\rm CNM}$ values, calculated by dividing the upper limits to $\Sigma \rm N(H\textsc{i}, CNM)_{20}$ by the total isothermal column density (Chengalur et al.\,2013), are all strong upper limits. 

\section{Comparison with synthetic observations}

Recently, Kim et al.\,(2014) used a hydrodynamical simulation of a Milky Way-like Galactic disk to construct synthetic H\textsc{i} absorption and emission profiles for thousands of sight lines probing the turbulent, multiphase ISM. They calculate ``observed" $T_s$, $T_k$, N(H\textsc{i}) and $f_{\rm CNM}$ for all lines of sight and found that the majority of sight lines with average LOS spin temperature below $200\rm\,K$ have a CNM fraction between 0.4 and 0.7 (98\% have $f_{\rm CNM}<0.7$; C-.G. Kim et al.\,2014 private communication). Whereas we measure several CNM fractions within this range, by inspection of Figure~\ref{f:nhlos} and Table~\ref{tab:los}, we have more measurements below $f_{\rm CNM}=0.3$ than predicted by the Kim et al.\,(2014) results. However, in agreement with HT03, Kim et al.\,(2014) also found many WNM-dominated sightlines with $f_{\rm CNM}=0$. 

Interestingly, Kim et al.\,(2014) found that synthetically-observed line-of-sight (harmonic mean) $T_s$ values agree within a factor of 1.5 with ``true" harmonic mean $T_s$ values computed from the simulation over the full range of integrated optical depth ($10^{-3}<\int \tau dv<10^{2}\rm\,km\,s^{-1}$). They argue that this implies that there is little overlap between CNM clouds along a line of sight within their $1\rm km\,s^{-1}$ velocity channels. However, by inspection of our absorption profiles and fits (e.g. Figure 4), it is clear that cloud complexes can overlap significantly along every real observed line of sight. 

\begin{figure*}
\begin{center}
\vspace{-5pt}
\includegraphics[width=0.9\textwidth]{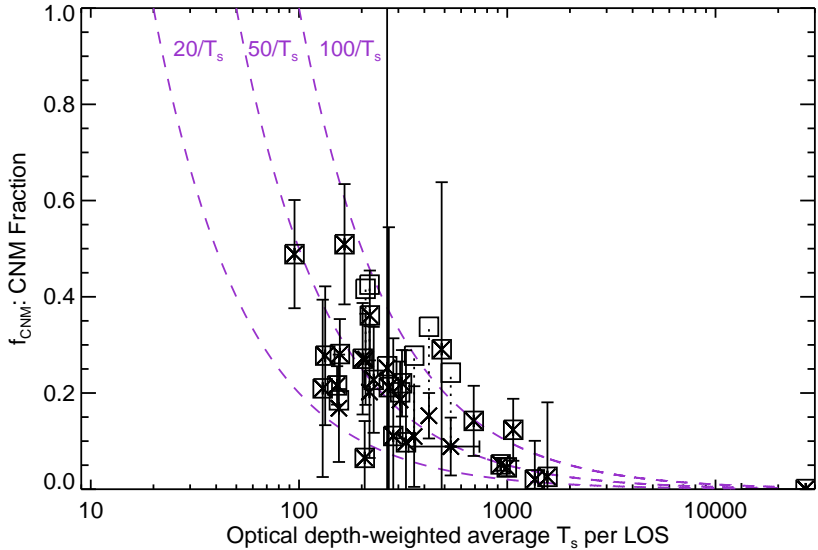}
\vspace{-300pt}
\caption{CNM fraction ($f_{\rm CNM}$) versus optical depth weighted average $T_s$ per LOS (Equation~\ref{e:tsobs}). Purple dashed lines indicate the observed trends in CNM fraction with LOS spin temperature from Kim et al.\,(2014) (their Figures 7a and 8a). Crosses indicate $f_{\rm CNM}$ calculated from Equation~(\ref{e:fcnm}) with a cutoff of $T_s=200\rm\,K$ to define the CNM, squares indicate $f_{\rm CNM}$ using a cutoff of $T_s=350\rm\,K$. Errors in $f_{\rm CNM}$ are shown (solid vertical lines) when they exceed the symbol size. Dotted lines connect crosses and squares when the difference between the two values exceed the symbol size.
}
\label{f:coldens2}
\end{center}
\end{figure*}

In Figure~\ref{f:coldens2} we plot CNM fraction versus the optical depth-weighted LOS spin temperature, $T_{s,\rm obs}$. This weighted spin temperature, $T_{s,\rm obs}$, is calculated from Equation~(15) of Kim et al.\,(2014), which was derived from the ``isothermal" column density estimate (Spitzer 1978, Dickey \& Benson 1982, Chengalur et al.\,2013, Kim et al.\,2014), and is given by,
\begin{equation}
T_{s, \rm obs}=\frac{\int \tau(v)~\frac{T_{\mathrm B}(v)}{1-e^{-\tau(v)}} dv}{\int \tau(v) dv}.
\label{e:tsobs}
\end{equation}

\noindent We display the error in the weighted $T_s$ values when they exceed the symbol size. The squares represent same quanitites with a cutoff for the CNM fraction at $T_s=350\rm\,K$ instead of $T_s=200\rm\,K$. This increases the value of $f_{\rm CNM}$ in only a few cases, because we only detect 6\% of all components in the $200<T_s<350\rm\,K$ regime by number. This shows that our choice of temperature cutoff does not strongly affect the observed trends in $f_{\rm CNM}$. 

Kim et al.\,(2014) plotted the same quantities (their Figures 7a and 8a) and found that $f_{\rm CNM}$, defined by a $T_k$ cutoff at $184\rm\,K$, varies with $T_{s,\rm obs}$ roughly as $f_{\rm CNM}\approx T_c/T_{s,\rm obs}$ for $50<T_c<100\rm\,K$. These $T_c/T_{s,\rm obs}$ curves are shown in Figure~\ref{f:coldens2} in dashed purple for $T_c=20,~50$ and $100\rm\,K$. Although we see a similar tight trend between CNM fraction and $T_{s, \rm obs}$, we find points below the $50/T_{s,\rm obs}$ curve, indicating again that we are measuring lower $f_{\rm CNM}$ and motivating us to extend the dashed curves down to $20/T_s$. Kim et al.\,(2014) possibly produced higher $f_{\rm CNM}$ because they did not include H\textsc{i}-H$_2$ chemistry, and therefore could have overestimated $f_{\rm CNM}$ if some cold, dense H\textsc{i} should have transitioned to H$_2$ form. This was also noticed by Stanimirovi\'c et al.\,(2014) in an H\textsc{i} emission and absorption line study of the Perseus molecular cloud at the Arecibo Observatory. 

Overall, we generally agree with the trend between CNM fraction and optical depth-weighted LOS $T_s$ ($T_{s, \rm obs}$) produced by the synthetic observations of the Kim et al.\,(2014) simulation, although our observations increase the scatter in the trend. 
The agreement is very encouraging considering that our CNM fractions rely on temperature calculation by Gaussian decomposition, whereas theirs are taken directly from their simulated lines of sight. 
Kim et al.\,(2014) derived an expression describing this trend based on several assumptions: $T_s=T_k$ in the CNM, the ``isothermal" column density equals the true column density (truest for low integrated $\tau$ sightlines), and the synthetically observed harmonic mean $T_s$ equals the true harmonic mean $T_s$ (where the agreement is best for $\tau<0.1$, above which there is scatter of up to 1.5). Finally, in computing $f_{\rm CNM}$, they assume that the CNM fraction is not ``extremely small" in order to express the $f_{\rm CNM}$ as a ratio of temperatures instead of column densities (see their Section 3.2, Equation (16), Kim et al.\,2014). This last assumption may be the most important to increasing the trend scatter, because we (and HT03) find many sightlines with both small $f_{\rm CNM}$ and small $T_{s, \rm obs}$.

The differences between our results are likely due in part to the fact that Kim et al.\,(2014) do not include magnetic fields or chemistry to map the transitions between H\textsc{i} and H$_2$, but the similarities found so far suggest that only small adjustments to the simulations may be necessary to match the observed $f_{\rm CNM}$. We also emphasize that the spatial resolution of a simulation is key for resolving the CNM and appropriately calculating quantities like the CNM fraction (e.g. Audit \& Hennebelle 2005). In the future we will apply our Gaussian fitting methods to large volumes of synthetic data from Kim et al.\,(2014) and other higher spatial-resolution numerical simulations using Autonomous Gaussian Decomposition (Lindner et al.\,2015). This will be essential to assessing the effect of our analysis techniques on these results and also understanding the biases of the numerical models.   

\section{Comparison of methods for estimating spin temperature from H\textsc{i} absorption and emission spectra}

Several independent methods are commonly used in the literature to measure $T_s$ from H\textsc{i} absorption and emission observations. The first is the method we use in this work, whereby we estimate a single $T_s$ value for each Gaussian component fitted to both absorption and emission spectra (i.e., solving Equation 1). This method relies on initial guesses for number of components and their shapes, but it is the only method based on a physical model. As is generally true for the $21\rm\,cm$ line, a Gaussian model is a good representation of the spectral line because the damping wings are insignificant. Although this assumption fails in some cases, shown in this study by the components with $T_{k,\rm max}>>T_s$ ($6\%$ have $T_{k,\rm max}>(T_s+3\sigma)$), the method is convenient and produces reasonable fits to both simple and complex profiles.

\subsection{One phase per channel $T_s$}

Assuming that the gas in each velocity channel is at a single temperature (the ``one-phase" assumption), dividing the brightness temperature spectrum ($T_{\rm B}(v)$) by the absorption profile ($1-e^{-\tau(v)}$) produces a spin temperature spectrum, $T_{s,O}(v)$, via the following equation,
\begin{equation}
T_{s,O}(v)=\frac{T_{\rm B}(v)}{1-e^{-\tau(v)}}.
\label{e:onephase}
\end{equation}

\noindent From this it is possible to measure a single $T_s$ value for every velocity channel. For example, Roy et al.\,(2013a) use this method to compute $T_{s,O}$ for each velocity channel above $3\sigma$ significance in absorption at $1\rm\,km\,s^{-1}$ resolution.  

\subsection{Line of sight harmonic mean $T_s$}

Assuming that there is one value of $T_s$ along a full line of sight, the ``LOS" harmonic-mean temperature, $T_{s,\rm L}$ from the absorption and emission profiles is given by,
\begin{equation}
T_{s,\rm L}= \frac{\int T_{\rm B}(v) dv}{\int (1-e^{-\tau(v)}) dv}.
\label{e:intLos}
\end{equation}

\noindent For example, Kanekar et al.\,(2011) computed $T_{s,\rm L}$ for the same data from Roy et al.\,(2013) to distinguish between CNM and WNM-dominated lines of sight.

\subsection{The Slope Method}

It is also possible to measure $T_s$ for individual gas components by estimating the slope or ``ridge line" of each linear feature in $T_B(v)$ vs. $1-e^{-\tau(v)}$ space (e.g. Mebold et al.\,1997, Dickey et al.\,2000). The advantage of this approach compared with Gaussian decomposition is that the solution can be determined directly from the data without fitting iterations or initial guesses. However, this method also requires the assumption that all gas in a given velocity range exists at a single temperature, as is also assumed in the one phase per channel and LOS methods. To improve the slope method parameter estimation, Dickey et al.\,(2003) developed a two-phase nonlinear least squares fitting technique to determine the $T_s$ slopes, which we will call the ``NLLSQ slope method". 

Dickey et al.\,(2003) carried out a detailed comparison of $T_s$ estimation methods, including Gaussian decomposition and the NLLSQ slope method, and found that in cases of isolated, unblended CNM components, these two methods agree reasonably. In another comparison study, HT03 computed $T_s$ for a sample of their sources using a by-eye version of the slope method and similarly concluded that the two methods agree in the case of unblended CNM components, even if there are blended, weaker WNM components present. Dickey et al.\,(2003) re-processed all 79 HT03 emission and absorption pairs using the NLLSQ slope method and found a median spin temperature per components of $T_s=31\rm\,K$, which is lower than the median value per fitted Gaussian functions by HT03 of $T_s=48\rm\,K$. This suggests that there may be systematic differences between the two methods. 

In the comparisons made by Dickey et al.\,(2003) and HT03, the $T_s$ values from the NLSSQ slope method tend to be lower than the Gaussian-derived $T_s$. This is due to the fact that spectral features in H\textsc{i} emission and absorption profiles do not always have the same shape. If channels in emission corresponding to narrow absorption have similar brightness temperature as channels with no corresponding absorption, the slope of the opacity feature will be small. HT03 note that this is because component emission is determined by the product of both optical depth and $T_s$, so that broad, warm components dominate emission and narrow, cold components dominate only in absorption. We will conduct a detailed comparison of the Gaussian and NLLSQ $T_s$ derivation methods upon completion of the 21-SPONGE data collection. In addtion, we plan to use synthetic spectra from numerical simulations, where the input $T_s$ values are known, to assess and quanitfy their associated biases. 

\subsection{Applying methods to 21-SPONGE data}

\begin{figure*}
\begin{center}
\vspace{-5pt}
\includegraphics[width=0.95\textwidth]{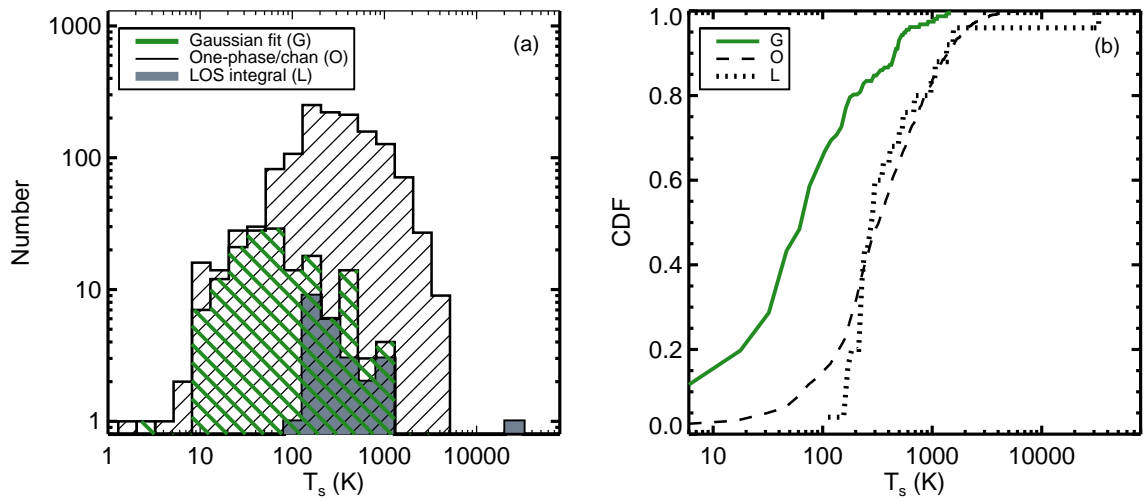}
\vspace{-340pt}
\caption{Comparison of temperature derivation methods for the 21-SPONGE survey. (a): Histograms of $T_s$ for: Gaussian decomposition of H\textsc{i} absorption and emission lines (thick green), ``one-phase" $T_s$ per 0.42 km/s channel (thin black, Equation~\ref{e:onephase}) and integrated $T_s$ per LOS (shaded gray, Equation~\ref{e:intLos}). (b): Cumulative distribution functions comparing the three temperature distributions from panel (a).
}
\label{f:methods}
\end{center}
\end{figure*}

From a preliminary analysis of 5 21-SPONGE sources (3C286, 3C225A, 3C225B, 3C237, and 3C298), we find that the NLLSQ slope method temperatures are mostly in the range $25<T_s<100\rm\,K$, which agrees well with the peak of our $T_s$ distribution (Figure~\ref{f:hist}a). In Figure~\ref{f:methods} we compare the Gaussian fit, one-phase per channel, and LOS harmonic mean methods by applying them all to the \NoOfSources{} 21-SPONGE emission/absorption spectral pairs. For the one-phase per channel method, for each source we used Equation~(\ref{e:onephase}) and included the $T_s$ values for all channels with absorption signals above the $3\sigma$ level. In Figure~\ref{f:methods}a, we plot histograms of $T_s$ for (1) the Gaussian decomposition method (thick green, ``G") (2) the one-phase per channel method (thin black, ``O") (3) the LOS harmonic mean method (shaded gray, ``L"). In Figure~\ref{f:methods}b we display cumulative distribution functions for the three $T_s$ methods.

Based on the CDFs of the three $T_s$ distributions in Figure~\ref{f:methods}b, the one-phase per channel and LOS-integrated methods agree well $(\mathrm{K}$-$\mathrm{S}=0.18, \mathrm{p}=0.35)$. Both methods rely on the assumption that there is a single-phase medium in each measurement interval (either velocity channel or full line of sight), which over-simplifies the reality that the ISM is a complex, multi-phase mixture of gas at many temperatures. The one-phase assumption overestimates $T_s$, because within each velocity channel (and line of sight) narrow components at lower temperatures can cumulatively produce a stronger absorption or emission signal, thereby biasing the average $T_s$ measurement to high values. As previously discussed by Dickey et al.\,(2000) and Dickey et al.\,(2003), the NLLSQ slope and Gaussian methods provide systematically lower values of $T_s$ than the single-phase approximations. This is supported by strong disagreements in Figure~\ref{f:methods} (p=0.00 between both G and O, and G and L).

The median values of the Gaussian, one-phase per channel and LOS methods are $T_{s, \rm G}=77\rm\,K$, $T_{s, \rm O}=340\rm\,K$ and $T_{s, \rm L}=290\rm\,K$. Wolfire et al.\,(2003) model the multi-phase ISM with detailed analytical prescriptions for heating and cooling, and show that the CNM should have a spin temperature of $40\le T_s\le200\rm\,K$. The Gaussian fit method best reproduces this predicted range, although the $T_s$ values in Figure~\ref{f:methods} also contain components associated with thermally unstable gas in addition to the dominant CNM. Furthermore, Figure~\ref{f:methods} shows that the Gaussian fit method is the most successful in reproducing the analytical prediction for the lower limit to CNM $T_s$. The LOS method in particular misses a large fraction of the CNM $T_s$ distribution, as it predicts all gas should have $T_s>200\rm\,K$. As discussed by Dickey et al.\,(2009), the $T_s$ single-phase (i.e. O and L) $T_s$ methods provide estimates of the mean temperature or the mixture of CNM and WNM along the line of sight, rather than physical temperatures of individual structures.

\section{Summary}

In this paper we present methods, data and preliminary results from the first half of 21-SPONGE, a large H\textsc{i} absorption survey at the Karl G. Jansky VLA. The absorption data are complemented by emission data from the Arecibo Observatory, whose $\sim3.5'$ beam at 21-cm provides the best available single-dish complement to the $\sim1''$ VLA beam so that we come closer by orders of magnitude than previous studies to sampling the same gas populations in absorption and emission. By obtaining extremely high-sensitivity VLA H\textsc{i} absorption lines, we detect signatures of both cold and warm ISM directly in absorption and, in combination with emission, directly determine the temperatures and column densities of individual components along the line of sight. We describe the observation and data reduction strategies and discuss details of line fitting and parameter estimation in this paper. Important initial results from the first half of the survey include: 

\begin{enumerate}
\item{We achieve median RMS noise in optical depth of $\sigma_{\tau}\sim9\times10^{-4}$ per $0.42\rm\,km\,s^{-1}$ channel over the \NoOfSources{} sightlines probed by 21-SPONGE so far. This allows us to directly probe absorption signals from both cold and warm H\textsc{i}. 
We are able to achieve this sensitivity in part by maximizing the signal-to-noise in our bandpass calibration solutions, which we accomplish by characterizing periodic structures within the solutions and combining calibration observations over time.}

\item{Following a careful Gaussian decomposition of all 31 H\textsc{i} absorption detections, we estimate $T_s$ and $\rm N(H\textsc{i})$ for \NoAbsComps{} Gaussian components. We find a maximum spin temperature per individual component of $T_s=1451\pm263\rm\,K$, which is more than a factor of 2 higher than the maximum Gaussian component spin temperature measured by the Millennium Arecibo H\textsc{i} Survey (Heiles \& Troland 2003; HT03). We are sensitive to lower $\rm N(H\textsc{i},\rm CNM)$ due to our better sensitivity in optical depth, and detect individual clouds with $\rm N(H\textsc{i},\rm CNM)$ as low as $3\pm1\times10^{16}\rm\,cm^{-2}$. }

\item{We compute total CNM column density, total WNM column density, total H\textsc{i} column density and CNM fraction along each of the \NoOfSources{} emission/absorption pair sightlines. We detect similar total CNM column densities, defined as the total column density of all Gaussian components per sightline with $T_s\leq200\rm\,K$, as HT03 and Roy et al.\,(2013). We find larger values of total ``warm gas" column density, defined as the total column density of all absorption-detected Gaussian components with $T_s>200\rm\,K$ and the components only detected in emission, due to the fact that we are sensitive to warmer gas in absorption than HT03. We find consistent CNM fractions ($f_{\rm CNM}$) with those found by HT03 and Roy et al.\,(2013), although the Roy et al.\,(2013) values are all upper limits.}

\item{We observe the same trend between CNM fraction and harmonic mean LOS spin temperature found by the synthetic-observation analysis of a hydrodynamic Galaxy simulation by Kim et al.\,(2014). Although we find lower CNM fractions (median 0.20) than the simulation predicts (98\% LOS with $0.4<f_{\rm CNM}<0.7$), the agreement in the trend is encouraging given that our calculations rely on Gaussian decomposition to estimate CNM fraction and theirs are taken directly from the simulation, suggesting that the Gaussian decomposition method is successful. }

\item{In a comparison of $T_s$ estimation methods, we find that one-phase or isothermal assumptions overestimate $T_s$ values relative to individual Gaussian fits. The $T_s$ distribution produced by multi-phase fits via Gaussian decomposition better reproduces theoretically expected lower limits for individual CNM clouds based on detailed heating and cooling analytical models.}

\end{enumerate}

\begin{acknowledgements}
We thank the referee for helpful comments and suggestions. 
This work was supported by the NSF Early Career Development
(CAREER) Award AST-1056780. C. E. M. acknowledges support by the National Science Foundation
Graduate Research Fellowship and the Wisconsin Space Grant Institution. S.S. thanks the Research
Corporation for Science Advancement for their support. 
We thank Nirupam Roy and Crystal Brogan for sharing absorption
spectra of the source 3C138 for comparison. We thank Eve Ostriker and Chang-Goo Kim 
for helpful discussions, and Ayesha Begum and Ben Klug for help with initial preparation and design for the survey.
The National Radio Astronomy Observatory is a facility of the National
Science Foundation operated under cooperative agreement by Associated
Universities, Inc. The Arecibo Observatory is operated by SRI International 
under a cooperative agreement with the National Science Foundation (AST-1100968), 
and in alliance with Ana G. M\'endez-Universidad Metropolitana, 
and the Universities Space Research Association. 

\end{acknowledgements}


\appendix

\section{Bandpass Calibration}

\subsection{59 kHz Ripple}

We find that all bandpass solutions obtained by 21-SPONGE contain a sinusoidal ``ripple" with a period of 59 kHz (or FWHM $\Delta v \sim$ 12 km/s at our velocity resolution) and amplitude of $\sim$0.0015. The period, amplitude and phase of this ripple are constant between array configurations and time of year. After investigating the stability of the ripple, we determined that it is caused by finite impulse response (FIR) filters in the WIDAR correlator applied prior to correlation which are used to shape the bandpass. The constant relative amplitude, phase and period of the ripple can all be explained in this context. 

\begin{figure}
\begin{center}
\includegraphics[width=0.49\textwidth]{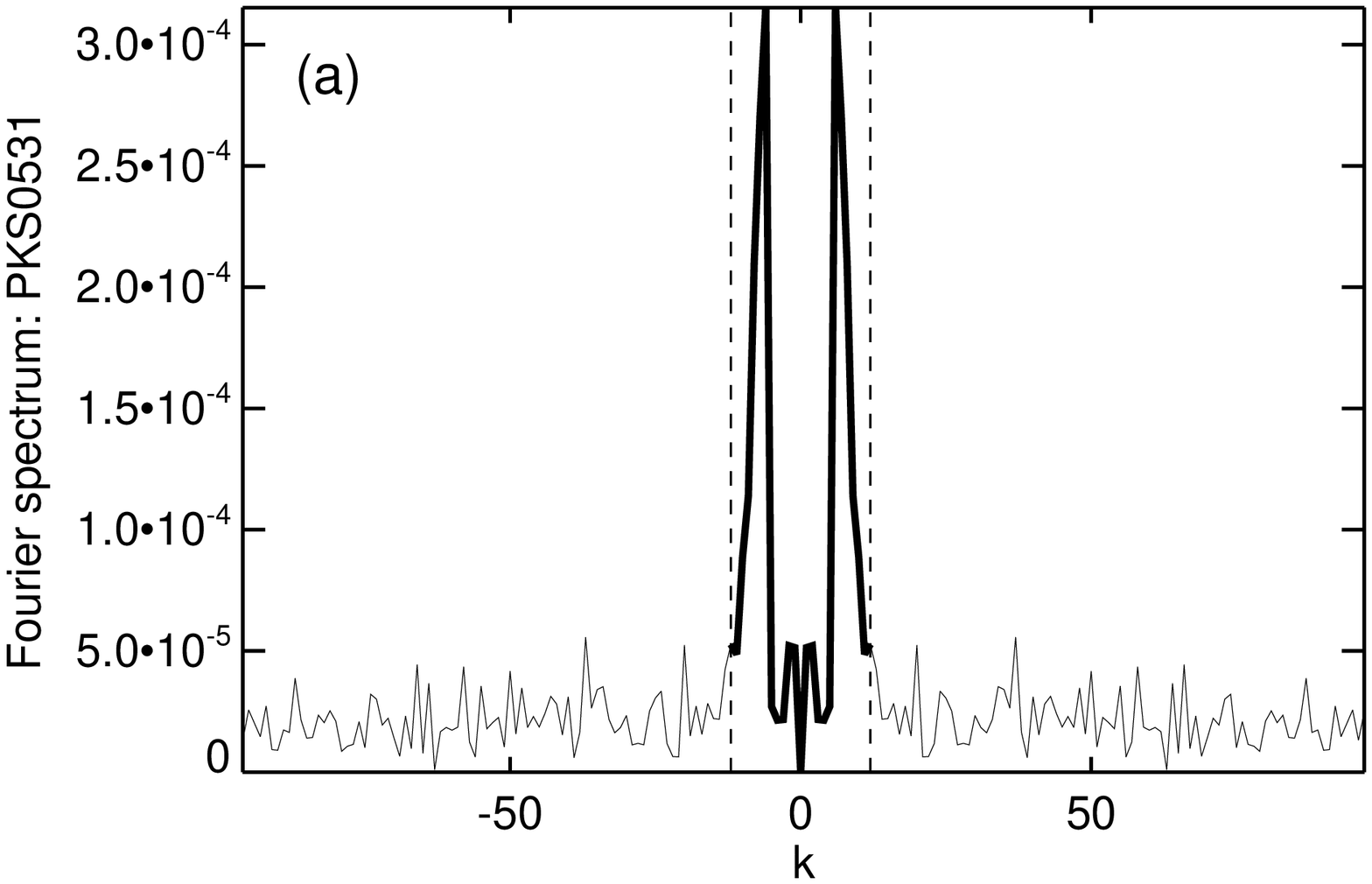}
\includegraphics[width=0.49\textwidth]{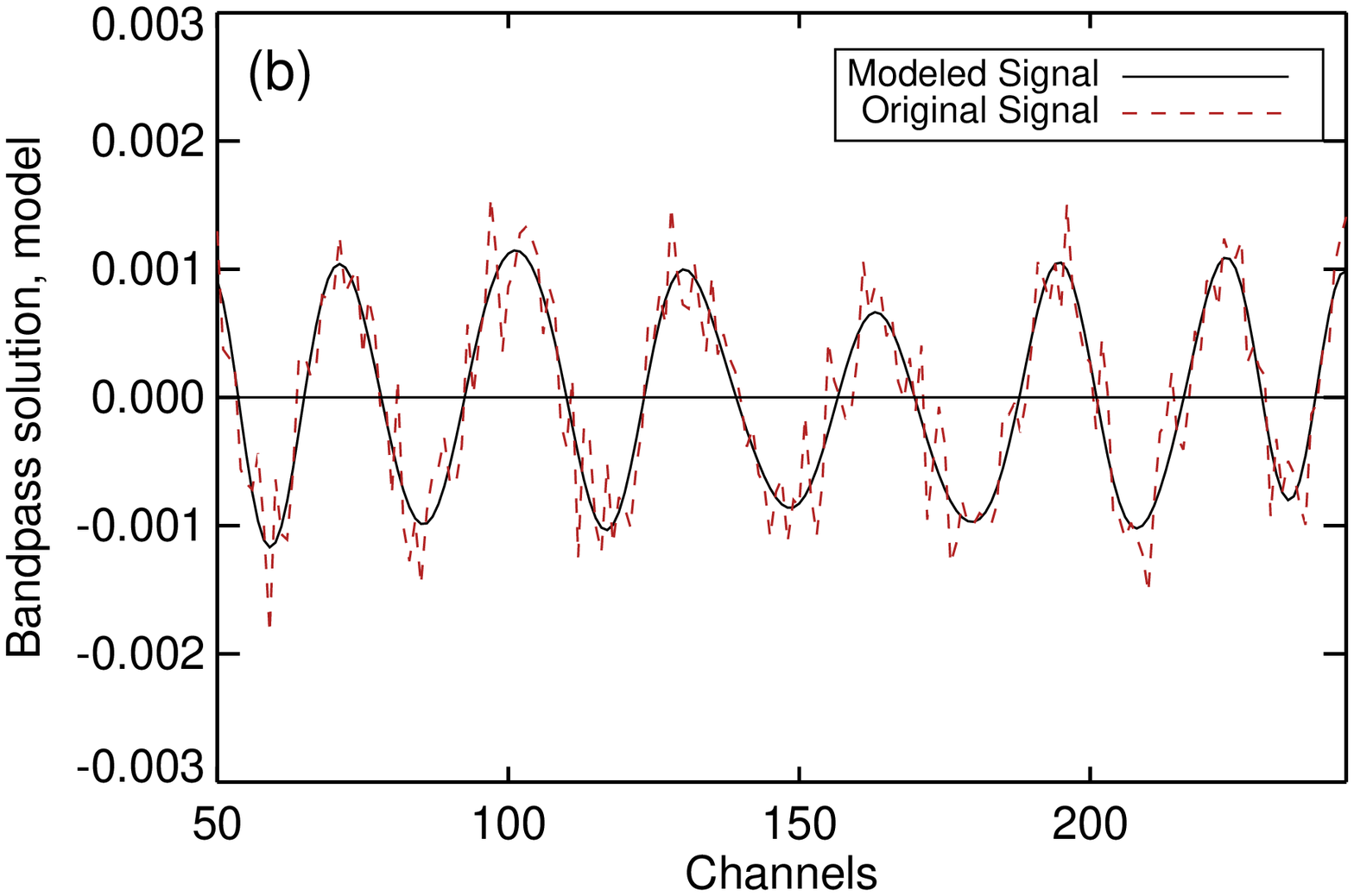}
\vspace{-120pt}
\caption{
(a): FT spectrum of a bandpass solution, with the boundaries of the applied step-function filter
overlaid as dashed lines. (b): Model of the bandpass solution (solid black),
computed as the inverse FT of the filter-convolved FT spectrum of the bandpass
solution. The bandpass solution is overlaid as the dashed red line.}
\label{fft1}
\end{center}
\end{figure}

To estimate the quantitative properties of the ripple, we first fit and remove the linear bandpass slope and zoom in on the central channels of the bandpass solution (channels 50-200 out of a total 256, shown in Figure~\ref{fft1}b). The linear slope does not affect our ability to detect wide absorption lines, and we are only interested in determining the parameters of the ripple and correcting for it. 

We then compute the Fourier transform (FT) spectrum of this flattened, zoomed solution (see Figure~\ref{fft1}a). Next, we multiply this FT spectrum with a simple step function filter to isolate the dominant periodic components. The boundaries of the filter are shown in Figure~\ref{fft1}a as vertical dashed lines. We then compute the inverse FT of the filter-multiplied FT spectrum (i.e., the thick black line in Figure~\ref{fft1}a) to model the periodic component. The modeled signal is displayed by the solid black line in Figure~\ref{fft1}b, along with the original slope-subtracted bandpass solution (dashed red line). 

\begin{figure}
\begin{center}
\includegraphics[width=0.49\textwidth]{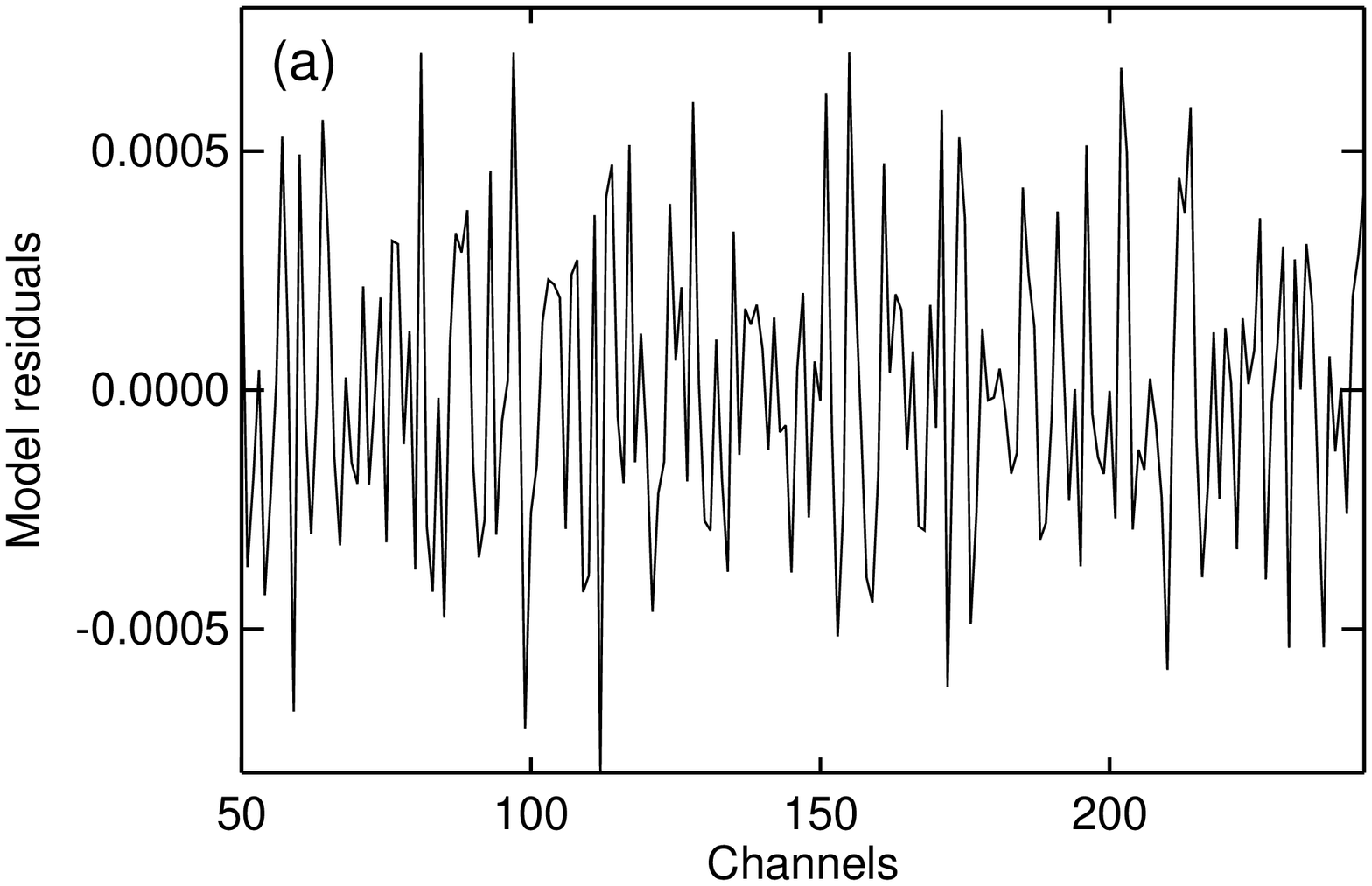}
\includegraphics[width=0.49\textwidth]{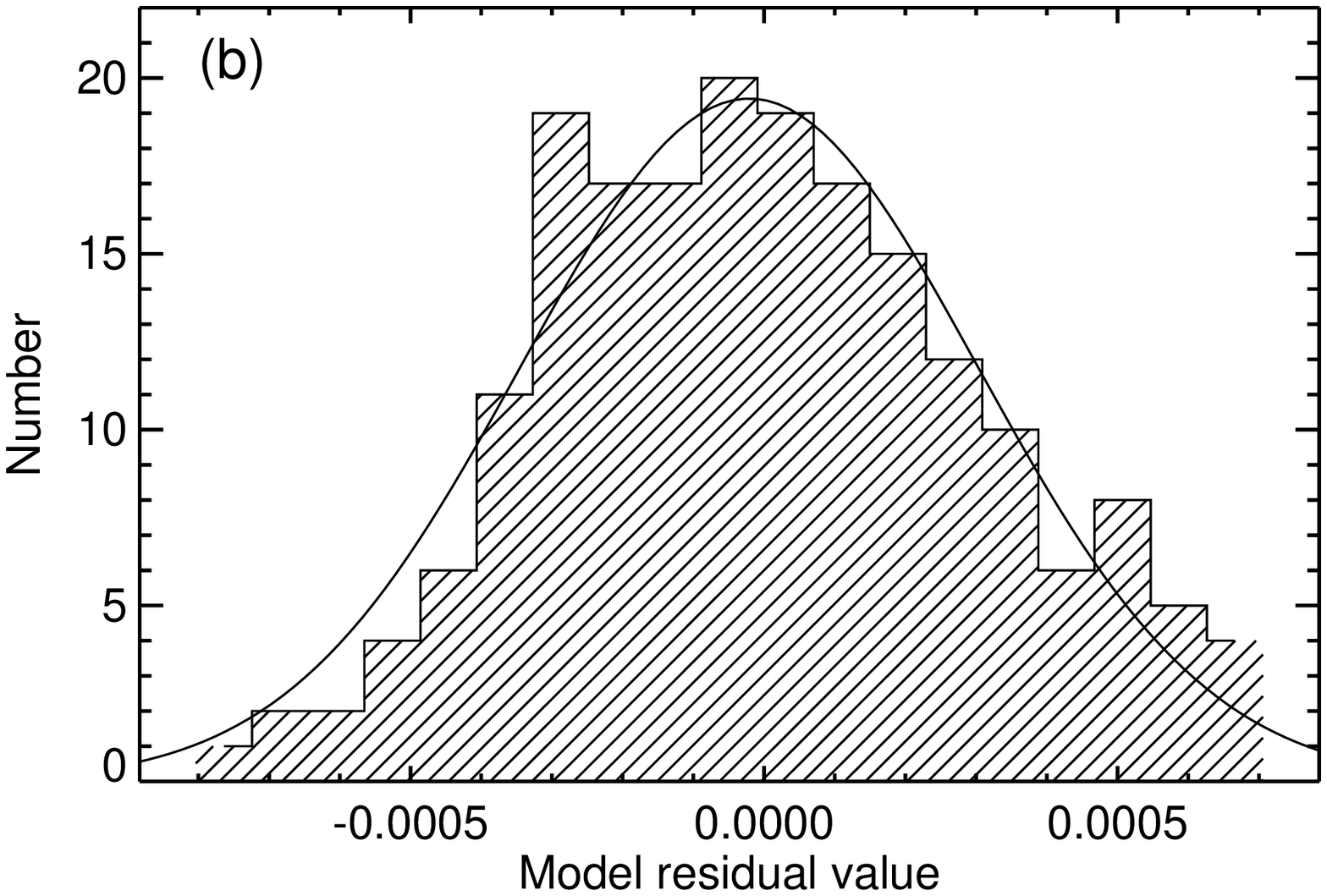}
\vspace{-120pt}
\caption{
(a): Residuals of the FT model to the bandpass solution displayed in
Figure~\ref{fft1} (i.e., bandpass solution minus model). (b): Histogram of the
residuals to check if they are well-represented by a Gaussian distribution, as
would be expected in the case of a successful model. A Gaussian model is
overlaid (solid black line) to illustrate this.}
\label{fft2}
\end{center}
\end{figure}

As a test of the success of the model, we compute the residuals by subtracting the model from the bandpass solution (see Figure~\ref{fft2}a). We then compute a histogram of these residuals to observe if they are well-represented by a Gaussian distribution, which we expect if the model is reasonable (see Figure~\ref{fft2}b). In this example, a K-S test between the residuals and a Gaussian distribution returns p=0.55, indicating that the noise is consistent with being drawn from a Gaussian distribution. From this, we can conclude that the model does a reasonably good job of isolating the dominant periodic components of the bandpass solution. We compute the RMS noise in these residuals as an indicator of the noise level in the bandpass solution that is not due to the periodic components. We record various parameters associated with the ripple fit in Table~\ref{info} for all bandpass solutions that we tested, including RMS noise in the model residuals, ripple amplitude, period and phase. The amplitude, period and phase of the ripple for each observation were calculated using a periodogram, following the methods of \citet{HB86}.

To test if the presence of the ripple is a function of particular antennas, we separated the bandpass solutions into three groups of about 9 antennas (i.e antennas 1-9, 10-18, 19-28). After fitting the solutions from these antenna subsets, we find that the ripple exists at a similar power as in the full solution in all cases. Therefore, we conclude that the effect is not antenna-based. 

\begin{table}[h!]
\caption{Bandpass Solution Parameters}
\vspace{5pt}
\centering
\label{info}
\begin{tabular}{lcccccc}
\hline
\hline
\small
\textbf{Information:} & & & \textbf{Fit Parameters:} & & & \\
Target Name   &  Obs. Date & Array Config. & RMS noise$^a$ & Amplitude$^b$ & Period$^b$ & Phase$^c$  \\
 & & & (per channel) & & (channels) & (channel) \\
\hline
3C120D1 & 03/11 & B &    0.00049 &     0.0014 &   30.1 &     120 \\
3C120D2 & 03/11 & B &    0.00047 &     0.0014 &   29.9 &     120 \\
3C286 & 05/11 & B &    0.00064 &     0.0016 &   29.9 &     122 \\
3C225BD1 & 05/11 & BnA &    0.00049 &     0.0015 &   30.0 &     121 \\
3C225BD2 & 05/11 & BnA &    0.00049 &     0.0015 &   30.3 &     121 \\
3C225BD3 & 05/11 & BnA &    0.00042 &     0.0013 &   30.1 &     120 \\
4C12.50 & 05/11 & BnA &    0.00048 &     0.0014 &   30.0 &     120 \\
3C345 & 06/11 & BnA-A &    0.00074 &     0.0017 &   30.1 &     120 \\
3C298 & 06/11 & BnA-A &    0.00057 &     0.0015 &   29.8 &     121 \\
4C32.44D1 & 05/11 & BnA-A &    0.00040 &     0.0014 &   29.9 &     122 \\
4C32.44D2 & 06/11 & BnA-A &    0.00063 &     0.0015 &   29.9 &     120 \\
4C32.44D3 & 06/11 & BnA-A &    0.00057 &     0.0015 &   29.8 &     122 \\
3C48 & 08/11 & A &    0.00048 &     0.0014 &   30.0 &     120 \\
4C16.09 & 09/11 & A &    0.00037 &     0.0014 &   30.0 &     120 \\
4C16.09.2 & 09/11 & A &    0.00031 &     0.0013 &   29.9 &     120 \\
3C133 & 09/11 & A-D &    0.00035 &     0.0014 &   30.1 &     120 \\
PKS0531 & 09/11 & A-D &    0.00031 &     0.0013 &   29.9 &     120 \\
3C111 & 04/12 & C &    0.00042 &     0.0014 &   29.9 &     120 \\
3C154D1 & 04/12 & C &    0.00057 &     0.0017 &   30.2 &     122 \\
3C154D2 & 04/12 & C &    0.00033 &     0.0013 &   29.9 &     120 \\
3C123 & 05/12 & CnB &    0.00041 &     0.0015 &   30.1 &     120 \\
3C138 & 05/12 & CnB &    0.00041 &     0.0014 &   29.9 &     122 \\
3C410 & 07/12 & B &    0.00057 &     0.0016 &   29.6 &     123 \\
\hline
\end{tabular}
\vskip 0.1 in
\footnotesize 
\raggedright
$^a$: the RMS noise calculated in the residuals of the bandpass solution
\emph{after} removing the ripple model. \\
$^b$: calculated using the periodogram analysis of the solution (i.e. these are the parameters of the dominant component) \\
$^c$: starting channel of a new period nearest to the center of the solution. 
\end{table}

\subsection{Time Averaging}

\begin{figure}[h!]
\begin{center}
\includegraphics[width=0.9\textwidth]{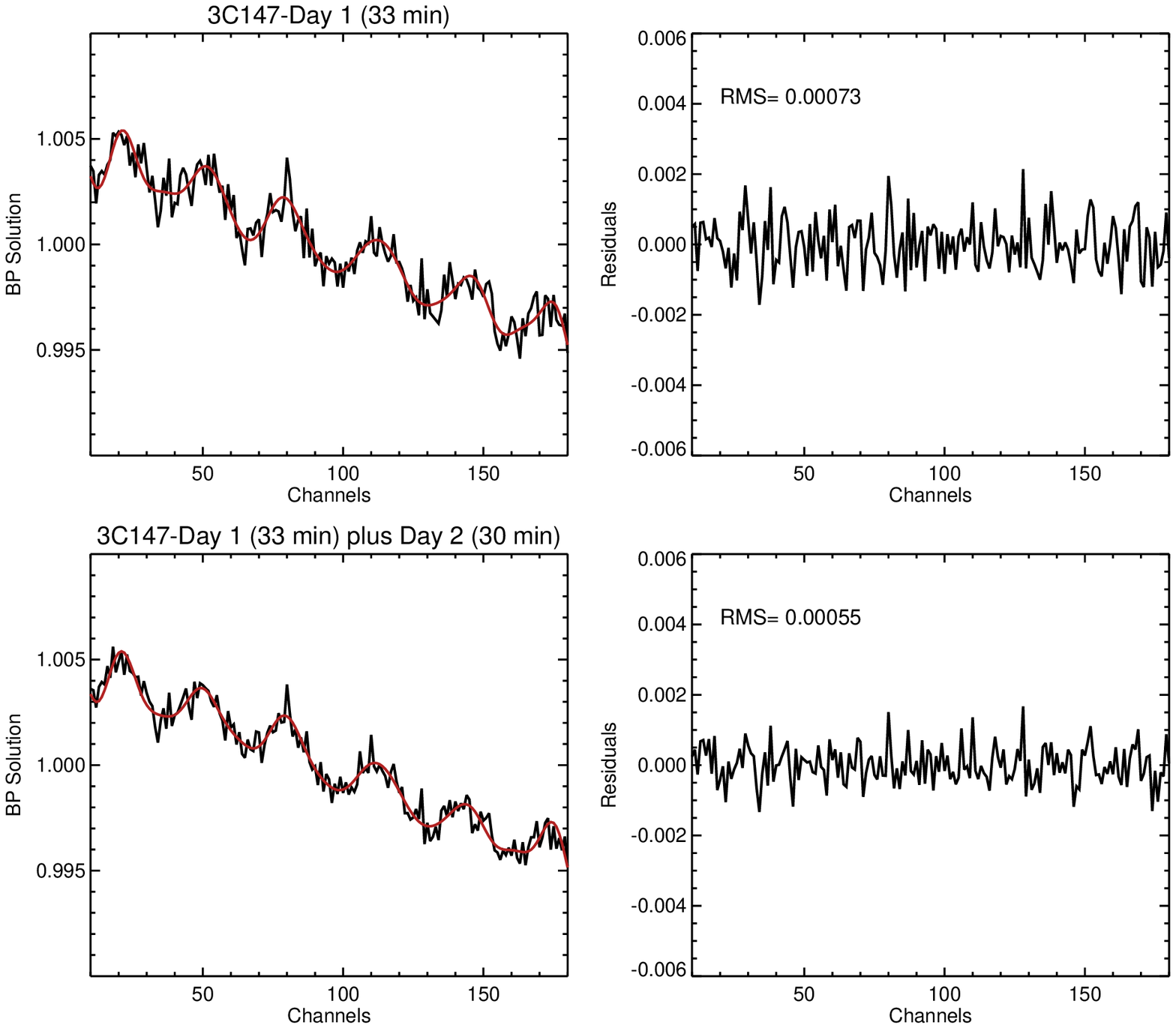}
\vspace{-190pt}
\caption{
An example of combining bandpass solutions. We fit the bandpass solutions (left) for the presence of a 59 kHz ripple, remove this model (shown in red) to produce the panels on the right, from which we measure the RMS noise in the solution. The top panels are from a single observation, and the bottom panels are the results of adding an observation 4 days later.}
\label{f:f1}
\end{center}
\end{figure}

To minimize noise levels in bandpass (BP) solutions, we combine bandpass observations acquired at different times. To compare noise levels in the BP solutions from the original observation and the combined observations, we first model and remove the 59 kHz ripple present in all solutions. Figure~\ref{f:f1} shows an example of the modeling process. A zoom-in onto a bandpass solution from one observation of 3C147 (33 min integration) is shown in the top left panel, with the ripple model overlaid in red. The top right panel displays the residuals  following the removal of the ripple model, from which the noise in the bandpass solution is computed.  The bottom panels are the same, after adding an additional observation of 3C147 from 4 days later (for a total of 33+30=63 min integration). 

We expect the noise in the bandpass solution to decline according to the factor of increase in integration time given by the ideal radiometer equation, or by $1/\sqrt{t_2/t_1}$. For example, from Figure 15, for an increase in integration time of $t_2/t_1=63/33= 1.9$, we expect an improvement in noise by a factor of $1/\sqrt{t_2/t_1}=0.72$. The improvement in noise is $0.00055/0.00073=0.75$. 

\begin{figure}[h!]
\begin{center}
\vspace{-10pt}
\includegraphics[width=0.8\textwidth]{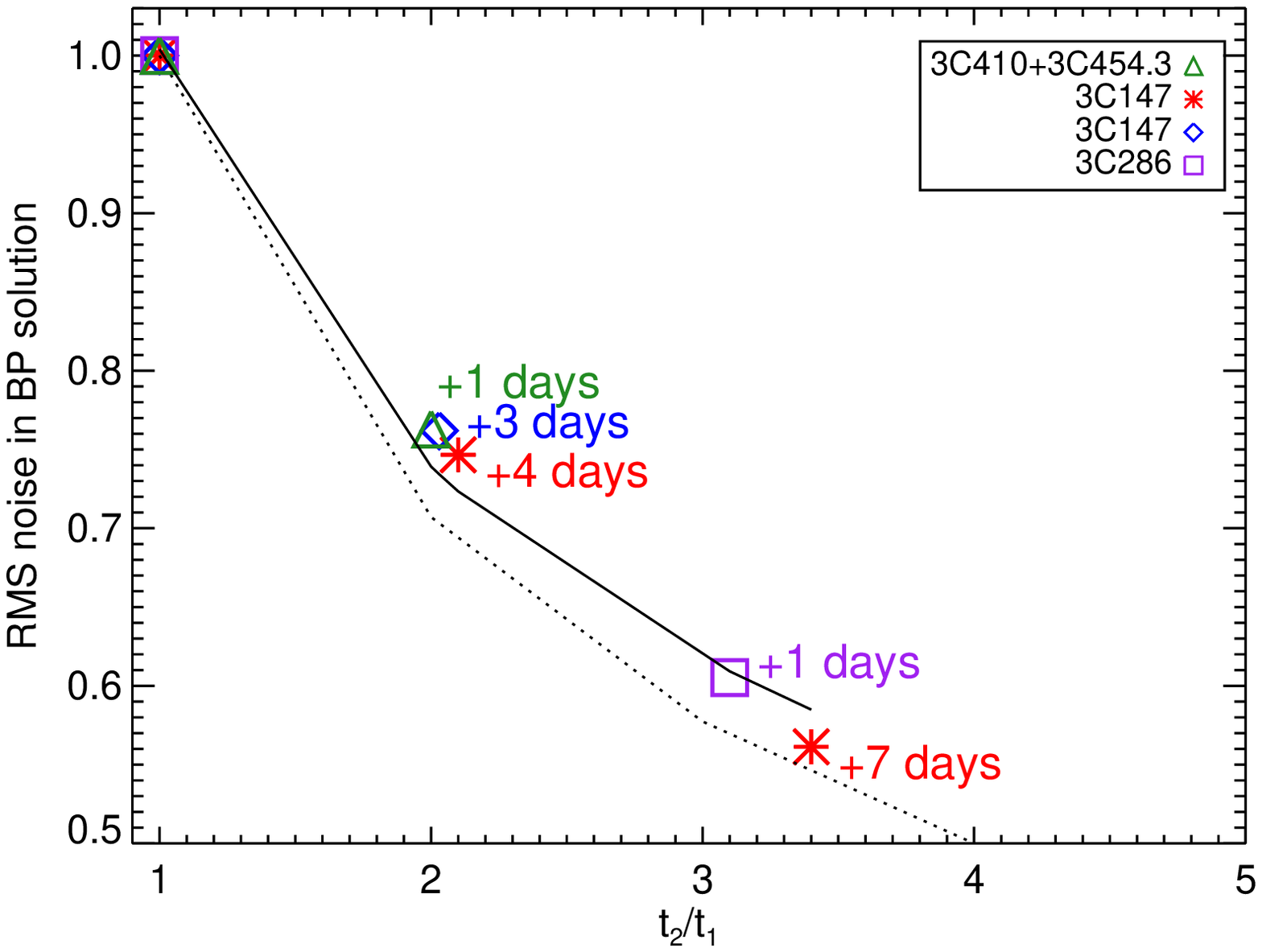}
\vspace{-120pt}
\caption{
The improvement in RMS noise in a bandpass solution (y-axis) by combining observations in time, thereby increasing the total integration time (x-axis). Symbols denote measurements, and the solid line is a linear fit to the data. The index of the fit is $-0.44\pm0.02$, compared to the theoretical expectation of an index equal to $-0.5$  (e.g. noise decrease by factor of $1/\sqrt{t_2/t_1}$, shown by the dotted line).}
\label{f:f2}
\end{center}
\end{figure}

In Figure~\ref{f:f2} we show the RMS noise in the bandpass solution (computed after removing the ripple) as a function of $t_2/t_1$ for several different observations (symbols). The solid line displays a fit to the data points, which is linear in log space with a slope of $-0.44\pm0.02$. This is nearly equal to the theoretically expected fit index of $-0.5$ (e.g. a noise decrease by factor of $1/\sqrt{t_2/t_1}$). This result is highly encouraging, as it demonstrates that VLA bandpass solutions at this resolution from different days are stable enough to be combined to increase the sensitivity of observations while significantly saving observing time, which is especially useful for very deep observations. 

In this process, we are able to combine observations of different sources. In Figure~\ref{f:f2}, the green triangles are from a trial combining observations of the sources 3C410 ($S_{1.4\rm\,GHz}=10\rm\,Jy$) and 3C454.3 ($S_{1.4\rm\,GHz}=11\rm\,Jy$) as bandpass calibrators. The other three cases are combinations of observations of the same bandpass calibrators, 3C147 ($S_{1.4\rm\,GHz}=23\rm\,Jy$) and 3C286 ($S_{1.4\rm\,GHz}=15\rm\,Jy$) between several days. These examples represent a common trend in our results. 

In later observations, we experimented with including additional 0.5 MHz subbands spaced $\pm 1.5,\rm\,3\,and 4.5$ MHz from the H\textsc{i} line. We found that combining these subbands does not significantly improve the noise in the bandpass solution.


\bibliographystyle{apj}

\label{lastpage}
\end{document}